# Design of Novel 3D SERS Probes with Drastically Improved Detection Limit by Maximizing SPP – Based Multiple Coupling Effects

*Yi Tian,†,‡ Hanfu Wang,† Lanqin Yan,† Xianfeng Zhang,† Attia Falak,†,‡ Yanjun Guo,† Peipei Chen,*,† Fengliang Dong,*,† Lianfeng Sun,*,†,‡ and Weiguo Chu*,†,‡*

† CAS Key Laboratory for Nanosystems and Hierachical Fabrication, Nanofabrication Laboratory, CAS Center for Excellence in Nanoscience, National Center for Nanoscience and Technology, Beijing 100190, P. R. China

‡ University of Chinese Academy of Sciences, Beijing 100049, P. R. China

**Corresponding Author**

* E-mail: wgchu@nanoctr.cn.  * E-mail: chenpp@nanoctr.cn.  * E-mail: dongfl@nanoctr.cn.

* E-mail: slf@nanoctr.cn.

ABSTRACT: Quantifying formidable multiple coupling effects involved in Surface-enhanced Raman scattering (SERS) is a prerequisite for accurate design of SERS probes with superior detection limit and uniformity which are the targets for trace substance detection. Here, combining theory and experiments on novel 3D periodic Au/SiO$_2$ hybrid nanogrids, we successfully develop a generalized methodology of accurately designing high performance SERS probes. Structural parameters and symmetry, Au roughness, and polarization are quantitatively correlated to intrinsic electromagnetic field (EMF) enhancements from surface plasmon polariton (SPP), localized surface plasmon resonance (LSPR), optical standing wave and their couplings theoretically, which is experimentally verified. The hexagonal SERS probes optimized by the methodology successfully detect $5\times10^{-11}$ M Hg ions in water, and $2.5\times10^{-11}$ M R6G with 40 times improvement of detection limit, an enhancement factor of $3.4\times10^8$ and uniformity of 5.56%, which results from the extra Au roughness - independent 144% contribution of LSPR effects excited by SPP interference waves as secondary sources, beyond the conventional recognition. This study opens up a pioneering way not only for providing the generalized design principles of SERS probe structures with high performance but for accurately designing their structures with particular purposes such as greatly improved detection limit and uniformity which are very significant for trace substance detection.

KEYWORDS: intrinsic electromagnetic field enhancement, localized surface plasmon resonance, surface plasmon polariton, SERS probe design, trace substance detection

Surface-enhanced Raman scattering (SERS) is a very powerful technique for molecule detection which has extensive applications in chemical, biological, and environmental fields.[1-4] The enhancement of Raman signals is primarily dependent on the electromagnetic field (EMF) produced on plasmonic substrates in which the achievable maximum intrinsic EMF determines the detection limit of SERS probes, being very significant especially for trace or single molecule detection.[5,6] Increasing the number of hot spots such as creating rich nanogaps between nanoparticles as possible is well recognized as a popular way to boost Raman intensity, which however wouldn't promote the detection limit normally.[7-9] So far, the enhancement of intrinsic EMF at hot spots capable of improving the detection resolution is achieved predominantly by reducing the gaps at hot spots and/or taking appropriate nanostructures.[10-13] Detection uniformity is another key performance parameter, usually determined by the distributions of hot spots and the structures of SERS probes.[10,12] Appropriate design of SERS probe structures would undoubtedly promote the detection uniformity, which is rarely reported.

Localized surface plasmon resonance (LSPR) can enhance the localized EMF greatly at hot spots,[14-16] while surface plasmon polaritons (SPPs) are also considered to contribute to the enhancement of EMF by coupling with LSPR and/or themselves upon propagating along metal/dielectric interfaces.[17,18] For the structure of metal nanoparticles / dielectric layers / metal films, part of LSPR energy can be transferred to SPPs while the EMF of interparticles is enhanced by the energy transfer between LSPR and SPPs by virtue of the underlying surface plasmon.[19-21] Couplings between LSPR and SPPs were also observed in some special structures, such as cavity-based arrays,[22] nanodish arrays,[23] nanopillar arrays,[24] hierarchical silver substrates,[25] 3D multi-branched nanostructures,[26] and gold bowtie nanoantenna[27] to enhance EMF through different mechanisms as well. To the best of our knowledge, so far the multiple coupling effects above are simply elucidated qualitatively instead of quantitatively owing to their complexities.[19,22,24,28] Quantitative studies of the contributions of individual coupling effects to SERS by enhancing the EMF of hot spots favor better understanding the properties of SPPs and LSPR, and make it possible to design high performance SERS probes with higher resolution theoretically. Fortunately, state of the art electron-beam lithography (EBL) with superior controllability, high degree of freedom, excellent uniformity of fabricated structures, and high precision and accuracy with a few nanometers[29,30] is qualified to realize the fabrication of diversified structures for quantitative studies as desired.

Herein, we developed novel SERS substrates by electron beam evaporating Au on 3D periodic $SiO_2$ frameworks with different symmetries and geometric dimensions to form $Au/SiO_2$ hybrid structures for quantitative studies of multiple coupling effects involved in intrinsic SERS such as SPP wave interference, incident light standing wave effects, LSPR effects and their coupling effects. By combining finite - difference time - domain (FDTD) calculations and mathematical analyses, geometric dimension, Au roughness, and polarization dependences of multiple effects for different nanogrids were established. Theory and SERS experiments agree very well, and thus a generalized methodology is proposed for design of SERS probe structures. The hexagonal SERS substrates optimized to maximize the strongest intrinsic electric field (EF) using the methodology experimentally achieve 40 times improvement of the detection limit for Rhodamine 6G (R6G) molecules with an SERS enhancement factor of $3.4\times10^8$ and uniformity of 5.56% (relative standard deviation, RSD). The methodology can give not only the generalized design principles of SERS probe structures but also their dimensions for performance optimization. Therefore, the robust approach of quantifying the multiple coupling effects to maximize the intrinsic EMF enhancements opens a new avenue to allow for accurate design of novel SERS probe structures with high performance, especially detection limit and uniformity, by combining state of the art nanofabrication technologies.

RESULTS AND DISCUSSION

We designed and fabricated triangular (*t*), square (*s*) and hexagonal (*h*) 3D periodic Au decorated silicon oxide hybrid nanogrids with various grid lengths $L_p$ and heights for exploration of symmetry and geometry dependences of SERS effect. Here, the sample of 36 nm thick Au deposited on hexagonal $SiO_2$ nanogrids with a grid length of 200 nm and a height of 198 nm is labeled by 36 nm Au/198 nm $SiO_2$_*h*200. All the abbreviations here are illustrated in Table S1. The defined geometric parameters of nanogrids, their fabrication and scanning electron microscopy (SEM) images are shown in Figure 1a and Figures S1-S4, respectively[31,32]. Figures S3 and S4 show that the interconnected $SiO_2$ nanogrids are rigid enough to stand firm even for an aspect ratio of sidewalls as high as 22 (9 nm wide and 198 nm high). The roughness of sidewalls tends to increase initially and decrease then with the increased Au thickness and/or height of sidewalls due to the lateral growth. Average 13 nm large Au nanoparticles with gaps and about average 7 nm thickness on each side, and their size distributions were statistically obtained from SEM images of 36 nm Au/198 nm $SiO_2$ nanogrids and fallen 36 nm Au / parallel $SiO_2$ nanowalls (for better SEM observations of Au nanoparticles on sidewalls, the 36 nm thick Au evaporated on 198 nm high $SiO_2$ nanowalls were fabricated which are easily fallen), as revealed by Figure 1b. Height dependences of sidewall width and cross sectional Au area were experimentally established to obtain the density of Au nanoparticles for FDTD calculations (Section S1, Figure S5 and Table S2). The larger the thickness of Au is, the more rapidly the Au area increases with height, implying the more effective deposition of Au on the more rough sidewalls.

The Raman intensity of the peak at 1360 cm$^{-1}$ for R6G is recorded to investigate SERS effects here, and its height dependences for 36 nm Au/SiO$_2$ nanogrids_$t$200, _$s$200 and _$h$200 with the $x$-polarized light are presented in Figure 1c. The intensity is found to increase more rapidly for the higher SiO$_2$ nanogrids, which probably results from the increase of sidewall roughness and/or extra effects with height. For both 27 and 36 nm thicknesses the intensity increases continuously with height whereas for 9 and 18 nm the intensity maxima are observed which shift to the larger height side for thicker Au, as shown in Figure S6. Furthermore, the heights which the maximum intensity corresponds to are just the same for triangular, square and hexagonal for 9 and 18 nm, respectively. This reveals that the intensity maximum is independent of structural geometry. For 9, 18 and 27 nm thicknesses the intensity for the same height decreases as $t$200 > $s$200 > $h$200 whereas for 36 nm the intensity decrease sequence changes as the height range changes, i. e. the intensity decreases still as $t$200 > $s$200 > $h$200 for those heights lower than 163 nm but the intensity decreases as $h$200 > $s$200 > $t$200 for the height of 198 nm. For the height of zero (the reference sample without sidewalls) their intensities are low enough to be negligible, as shown in Figure 1c. By assuming reasonably that different SERS effects for different nanogrids arise predominantly from different geometries, intrinsic SERS effects can be evaluated using the SERS intensities normalized by their corresponding ratios of sidewall to bottom area, $S_s/S_b$, basically proportional to the density of hot spots (the number of hot spots per unit area), as shown in Figure 1d-1f (the density of hot spots changes as $t200 > s200 > h200$, in accord with the intensity change sequence for the heights lower than 163 nm above). The normalized intensities of $t200$, $s200$ and $h200$ for 27 and 36 nm Au increase continuously with $S_s/S_b$ (height) whereas the maximum intensities are seen for 18 nm at 163 nm height (Figure S6). Interestingly, the normalized intensity for $h200$ (the lowest density of hot spots) with 36 nm thick Au increases most rapidly with $S_s/S_b$ whereas that for $t200$ (the highest density of hot spots) increases most slowly. This reveals that the strongest intrinsic SERS effect for $h200$ especially for larger heights cannot be attributed to the rough sidewalls simply but extra effects produced by the hexagonal geometry. Therefore, we need to explore the underlying mechanisms of the intrinsic SERS effects by deriving quantitative dependences of multiple coupling effects on influencing factors for different nanogrids.

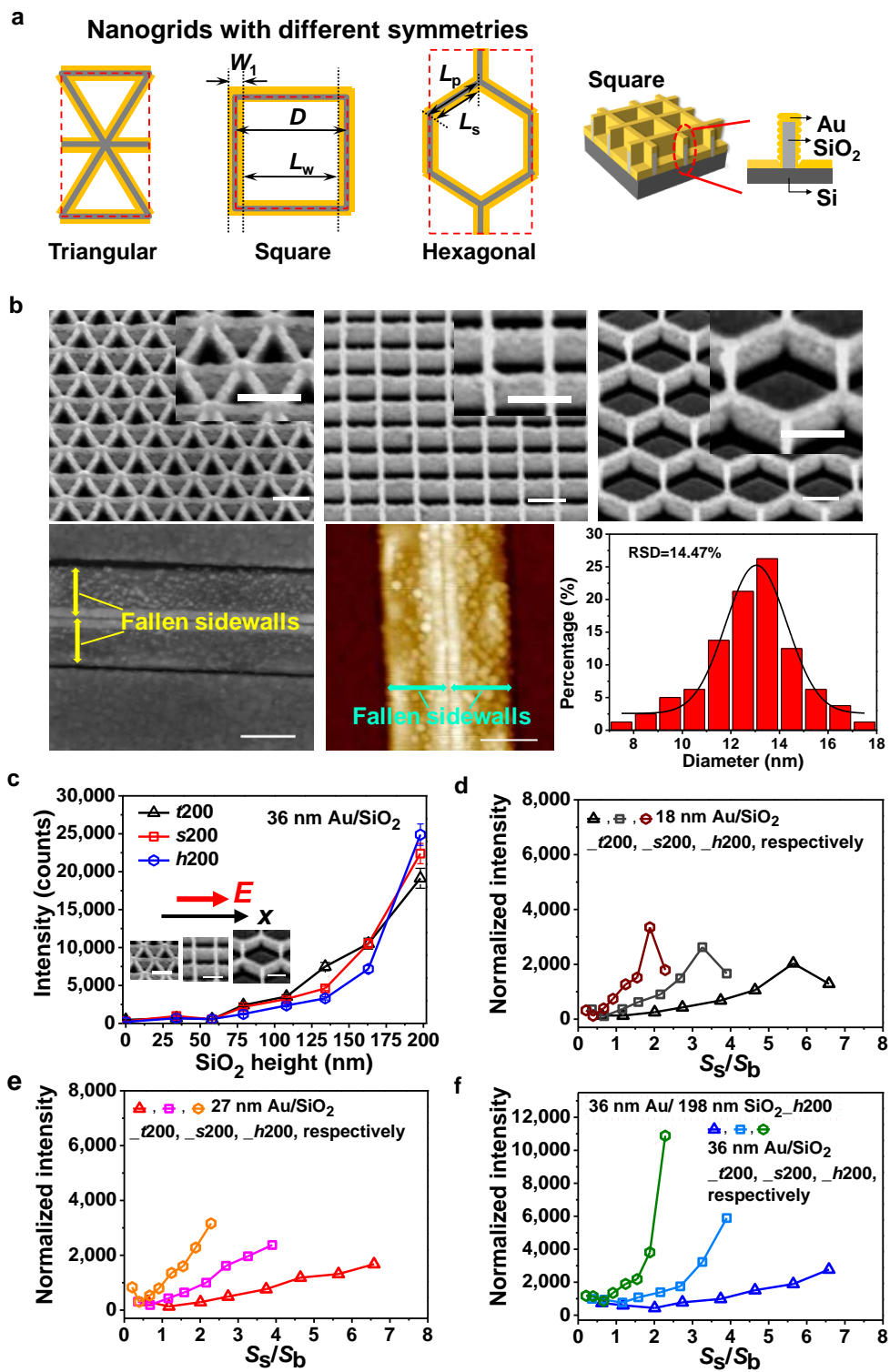

**Figure 1.** Structures of Au/SiO$_2$ nanogrids and experimental SERS intensities. (a) Schematics of triangular, square and hexagonal Au/SiO$_2$ periodic nanogrids. The unit cells are indicated by the red dashed rectangles. Spacing, width and center distance of sidewalls $L_w$, $W_1$ and $D$,

respectively, are defined in the schematics, and thus $L_w = D - W_1$. Grid length $L_p$ and sidewall length $L_s$ are also defined in the schematics, and thus $L_s \approx L_p - W_1$. (b) Tilt SEM images of triangular, square and hexagonal 36 nm Au/198 nm SiO$_2$ nanogrids with grid length $L_p$ of 200 nm and their corresponding enlarged images in the insets, and SEM and AFM (the presence of broadening effects) images of fallen 36 nm Au/198 nm SiO$_2$ parallel nanowalls and their size distribution derived from these images. Scale bars: 200 nm. (c) Changes of experimental SERS intensities of triangular ($t$200), square ($s$200) and hexagonal ($h$200) 36 nm Au/SiO$_2$ nanogrids with SiO$_2$ height. (d) - (f) Normalized SERS intensities of 18, 27 and 36 nm Au/SiO$_2$ nanogrids _ $t$200, $s$200 and $h$200 by the ratio of sidewall ($S_s$) to bottom surface area ($S_b$) versus the ratio ($S_s/S_b$, height), respectively.

For rough Au on periodic SiO$_2$ nanogrids, SPP waves and their interference effects, LSPR effects, standing wave effects of incident light and coupling effects between LSPR and SPP are possibly involved, which are undoubtedly responsible for their SERS effects. SPP1 can be excited at the Au/air and Au/SiO$_2$ interfaces of sidewalls by a polarized light with TE and TM modes,[33] and SPP2 at the bottom Au/air interface with TM mode due to the periodic sidewalls,[34] as schematically illustrated in Figures 2a, 2c and Figure S7a. LSPR1 and LSPR2 can be excited at the gaps between bottom and sidewalls, and at those on rough sidewalls under the light illumination, respectively, as shown in Figure 3a. The bottom Au film can reflect the incident light to form the optical standing wave, which may lead to the fluctuating distribution of electric field. SPP1 wave at the sidewalls (SPP2 wave at the bottom) can interfere with itself to form the interference wave due to periodic structures which can further excite extra LSPR as secondary sources. Therefore, the incident light is taken as primary source, and thus the total intensity of LSPR is the sum of those excited by the primary and secondary sources.

SPP could interfere in a cavity due to multiple reflections, like FP resonance,[35,36] and their wavelengths at the Au/air and Au/SiO$_2$ interfaces are derived to be 603 and 389 nm for the excitation wavelength of 632.8 nm, respectively (see Section S2 and Figure S7). SPP1 wave interference effects for TE mode with $\alpha = 0°$ (for TM mode with $\alpha = \pi/2$, Figures S8 and S9a-b) can be derived by comparing the FDTD calculations on the bottom Au - free rough nanogrid and nanowall models in Figure 2a, as shown in Figure 2b. Clearly, $I_{\text{SPP1-interference}}$ for triangular, square and hexagonal nanogrids changes with sidewall length $L_s$ with the maxima observed which are all achieved at $L_s = 302$ nm. This is actually the total resulting interference effects of SPP1 waves at the Au/air and Au/SiO$_2$ interfaces with the maxima which are achieved at $L_s = 302$ and 195 nm (the integral multiple of corresponding half $\lambda_{\text{SPP1}}$, Figure S7b), respectively, similar to the geometrical conditions that the FP resonance occurs.[35,36] In addition, it can be seen that $I_{\text{SPP1-interference}}$ also changes with $\alpha$ for the model of one sidewall for triangular, square and hexagonal 36 nm Au/198 nm SiO$_2$ nanogrids with $L_s = 302$ nm, which can be well fitted by using the expression of $(I_{1,\pi/2}-I_{1,0})\sin^2\alpha+I_{1,0}$ ($I_{1,\pi/2}$ and $I_{1,0}$ are the interference intensities of SPP1 wave for $\alpha = \pi/2$ and 0, respectively) (Figure S9c).

Similarly, SPP2 wave interference effects can also be extracted by comparing the smooth periodic nanogrid (with SPP2 effects) and $x$-nanowall models (without SPP2 effects) in Figure 2c. The distributions of the squared EF intensities (i.e. light intensities) at sidewalls ($xz$ plane) for the squared periodic nanogrid model ($E^2\_s200$) and for the $x$-nanowall model ($E^2\_x200$) are shown in Figure 2d. The ratio of $E^2\_s200$ to $E^2\_x200$ equals $(1+I_{\text{SPP2-interference}})/1$ from which $I_{\text{SPP2-interference}}$ can be derived, as shown in Figure 2e. $I_{\text{SPP2-interference}}$ is found to oscillate with sidewall spacing $L_w$, as done by the FP resonance.[35,36] Height dependences of the electric field intensity mathematically derived (Section S3 and Figure S10) clearly show that the maximum normalized intensity can be achieved at the height of 228 nm (36% of the incident light wavelength), close to 198 nm, the maximum height of Au/SiO$_2$ nanogrids taken here (Figure S5c), which is caused by the optical standing wave effect of incident light. This reveals for the first time that the height of a SERS probe structure, 36% of the incident light wavelength employed, is adequate for the achievement of the maximum normalized EF intensity, and further increasing height would lead to a gradually decaying oscillation of normalized EF intensity. Height dependence of the average interference intensity of SPP2 wave for $L_w$=177 nm is quite weak, as shown in Figure 2f, which is attributed to the destructive interference of SPP2 occurring at $L_w$=177 nm, as Figure 2e displays. In sharp contrast, the average interference intensity of SPP2 wave is far more dependent on height for $L_w$ = 302 nm where the strongest FP resonance - like interference effect of SPP2 wave takes place, which is clearly revealed by Figure S11.

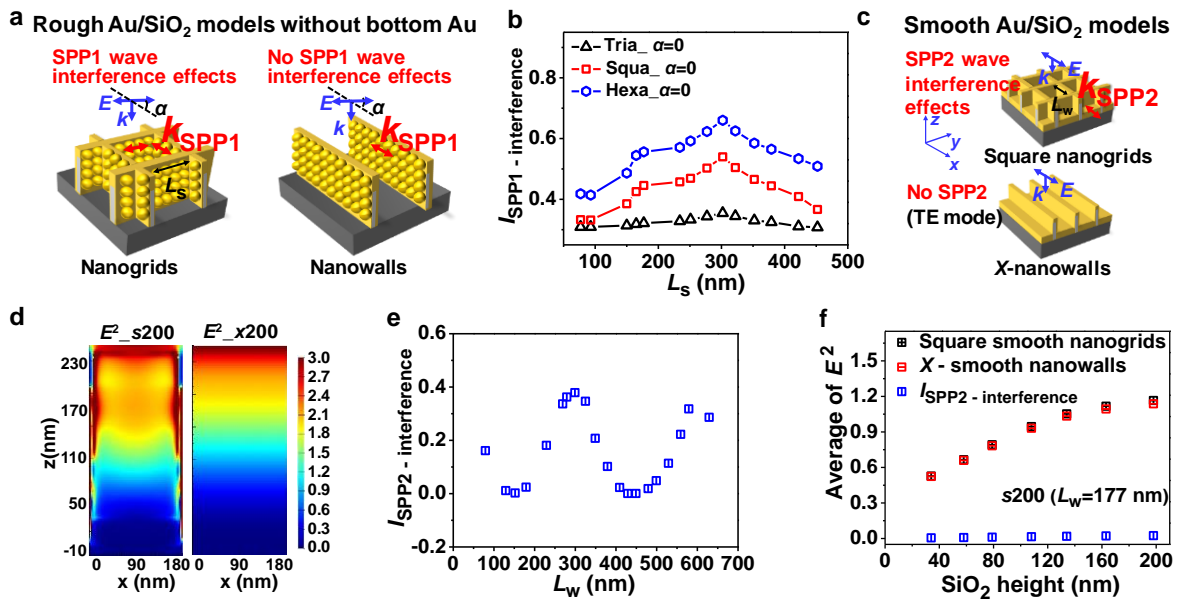

**Figure 2.** Geometric dimension dependences of SPP effects derived from FDTD calculations based on different 3D Au/SiO$_2$ models. (a) 3D models of the rough square Au/SiO$_2$ nanogrids and nanowalls without bottom Au. The comparison of the effects calculated from two models allows one to derive the SPP1 wave interference effect. (b) Calculated interference intensities of SPP1 wave excited at the rough sidewalls of triangular, square and hexagonal 36 nm Au/198 nm SiO$_2$ nanogrids with the polarization angles of $α = 0$ against the increased sidewall length $L_s$. (c) 3D models of the smooth square Au/SiO$_2$ nanogrids and *x*-nanowalls. Similar to the case in (a), the comparison of the effects produced by two models allows for the derivation of the SPP2 wave interference effect. (d) Calculated spatial distributions of the squared electric field intensity at surfaces (parallel to the *xz* plane) of the smooth square 36 nm Au/198 nm SiO$_2$ nanogrids_*s*200 and *x*-nanowalls_*x*200 for the *x*-polarized light. (e) Calculated interference intensities of SPP2 wave for the smooth square 36 nm Au/198 nm SiO$_2$ nanogrids with $L_w$. (f) Calculated averages of the squared electric field intensities at surfaces (parallel to the *xz* plane) of the smooth square 36 nm Au/SiO$_2$ nanogrids and *x*-nanowalls, with $D = 200$ nm (i.e. $L_w = 177$ nm, $L_w = D−W_1$ ($W_1$=9 nm+7×2 nm)), which reflects the incident light standing wave effect and interference intensities of SPP2 wave for the square nanogrids_*s*200 with SiO$_2$ height for the *x*-polarized light.

The geometric dimensions where the maximum interference effects of SPP1 and SPP2 waves at sidewalls occur have been derived above. Besides this, SERS intensity is also influenced greatly by the EMF enhancement from LSPR which is dependent on the size, gap and morphology of neighboring particles.[14-16] Here, we adopt semiellipsoids (hemispheres) for 27 and 36 nm (18 nm) thicknesses with $d < 2b$ (intersected), $d = 2b$ (tangent) and $d > 2b$ (separated) to describe the relationship between neighboring Au nanoparticles and further the change of sidewall roughness with height,[37] as shown in the left panel of Figure 3a. The radius / semi-major *b*, number *n*, center spacing *d*, of and the gap *g* ($g = d - 2b$) between hemispheres / semiellipsoids are derived from close inspection of SEM images, size distributions and calculations (Figure 1b, Figure S5 and Table S3). Figure 3b shows gap *g* dependences of the averaged EF enhancements $|E/E_0|^4$ by LSPR2 for 18, 27 and 36 nm Au/SiO$_2$ nanowalls_*D*200 with TE mode in the right panel of Figure 3a regardless of the coupling effects of SPP1 waves and of the optical standing wave with LSPR. The averaged $|E/E_0|^4$ reaches the maximum for $g \to 0$ and drops rapidly with $g$ away from zero (the smaller the absolute gap is, the larger the roughness of Au is for the same thickness). Here, it should be pointed out that quantum correction don't need to be taken into account upon calculations for the infinite small gap because the Au nanoparticles with different neighboring relationships always reside on a continuous Au film which allows to neglect the tunneling effect of electrons occurring among neighboring nanoparticles possibly.[38, 39] For the same $g$ the averaged $|E/E_0|^4$ increases more rapidly for the larger thickness because of the larger roughness arising from the lateral growth of more Au. Therefore, the particle models with different gaps proposed above can unambiguously describe the change of sidewall roughness with Au thickness. In addition, the size distributions of Au nanoparticles are normally inevitable in real cases. Here, our purpose is to describe the change of sidewall roughness correctly by using the different models of Au nanoparticles with single average sizes instead of the actual sizes with distributions.

With the parameters of Au nanoparticles obtained above, height dependences of LSPR effects for 36 nm Au / $SiO_2$ parallel rough nanowalls _ $D$200 for TM and TE mode in the right panel of Figure 3a are derived by FDTD calculations to be shown in Figure 3c. For the couplings of SPP1 and SPP2 with LSPR1 for TM mode, LSPR effects are almost independent of height with the maximum localized EF intensity of ~20 whereas for the coupling of SPP1 with LSPR2 for TE mode, LSPR effects are definitely height dependent with the maximum intensity of ~50 achieved at the height of 198 nm. The EF distributions for 18 and 27 nm thicknesses with various heights are shown in Figure S12. The maximum intensity including the coupling of SPP1 with LSPR2 for 18 nm thickness is observed at the height of 163 nm by considering the large roughness of Au, quite comparable to those for 27 and 36 nm thicknesses. This can well explain the experimental intensity maxima observed for triangular, square and hexagonal 18 nm Au/163 nm SiO$_2$ nanogrids shown in Figure 1d-1f. Localized EFs for 27 and 36 nm thicknesses change similarly with height but are weaker for 27 nm thickness. Therefore, the Au thickness of 36 nm and the SiO$_2$ height of 198 nm are revealed by calculations to be the best combination for EMF enhancement here, as shown by the experiments as well. The difference of the EF intensities at the gaps of Au nanoparticles for different heights arises mainly from the optical standing wave effect of incident light (Figure 2f and Figure S10).

Sidewall spacing $L_w$ dependences of the averaged $|E/E_0|^4$ from LSPR1 and LSPR2 at TM and TE mode for 36 nm Au/198 nm SiO$_2$ rough nanowalls are given in Figure 3d. The average $|E/E_0|^4$ from both LSPR1 and LSPR2 achieves the maxima at $L_w$ ~300 with far stronger LSPR2, and the maximum $|E/E_0|^4$ from LSPR2, i. e. the strongest intrinsic EF shows the maximum at $L_w$ ~300 nm for the rough nanowalls with TE mode. In addition, polarization dependences of EF enhancement for 36 nm Au/198 nm SiO$_2$ rough nanowalls with $L_w$ = 151 nm are also given in Figure 3e in which $L_w$ = 151 nm is taken to minimize the interference effect of SPP2 wave (Figure 2e). The EF enhancements from LSPR1 and LSPR2 increase and decrease with $α$, respectively. We give the change of the normalized $|E/E_0|^4$ from LSPR1 and LSPR2 against $α$ in Figure S13 which can be well fitted with $\sin^4\alpha$ and $\cos^4\alpha$, respectively. Therefore, the LSPR effects are influenced significantly by the polarization angle of $\alpha$ for the one dimensional nanowall model. However, this also implies that the influences would be probably smeared out for the structures with a high symmetry.

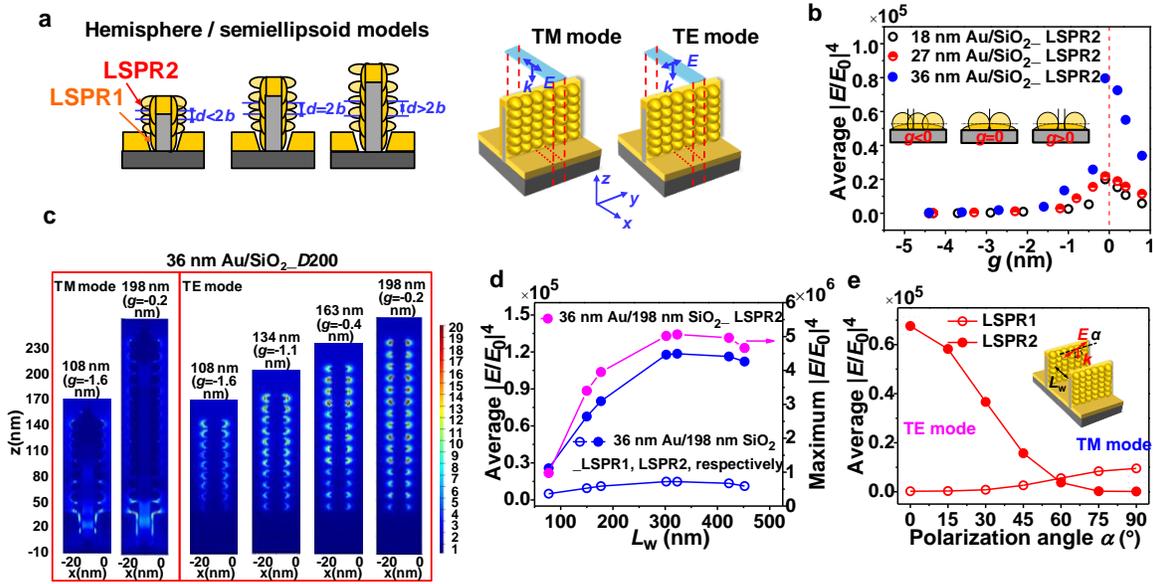

**Figure 3.** Calculated geometric dimension and polarization dependences of LSPR effects. (a) The models of hemisphere / semiellipsoid Au nanoparticles for the rough $y$-nanowalls with the increased height, accompanied by the initial increase and subsequent decrease in roughness, and the 3D rough $y$-nanowall models with TM and TE mode for the $x$ and $y$ polarization, respectively, for FDTD calculations. The unit cells are indicated by the red dashed lines. (b) Gap $g$ dependences of the averaged fourth power of local electric field intensities of the rough nanowalls with $D = 200$ nm for TE mode for 18, 27 and 36 nm thicknesses. (c) Calculated spatial distributions of the electric field intensities at the cross sections parallel to the $xz$ plane of the rough $y$-nanowalls with 200 nm center distance $D$ and 36 nm thickness for different heights with TM and TE modes. Semiellipsoid Au nanoparticles have the average semi-principal axes, $a=b=6.5$ nm, and $c=8.0$ nm. (d) Sidewall spacing $L_w$ dependences of the averaged $|E/E_0|^4$ of LSPR1 and LSPR2 with TM and TE mode, respectively, for the rough $y$-nanowalls with 36 nm Au/198 nm SiO$_2$, and the maximum $|E/E_0|^4$ of LSPR2 for these nanowalls for TE mode (right $y$-coordinate). (e) Calculated average $|E/E_0|^4$ of LSPR1 and LSPR2 versus polarization angle $\alpha$ for the rough 36 nm Au/198 nm SiO$_2$ nanowalls with $L_w = 151$ nm. The inset shows the model for FDTD calculations.

To generalize and extend the analysis method established based on the square nanogrid and parallel nanowall models to the triangular and hexagonal nanogrid structures, we reasonably take a triangular (square, hexagonal) nanogrid as consisting of three (two couples, three couples) nanowalls with 60° (90°, 120°) relative to each other, and decompose a linearly polarized incident light into two orthogonal components parallel and normal to each nanowall, i.e. TE and TM mode, respectively, which are schematically illustrated in Figure S14. Thus, the total average $|E/E_0|^4$ for triangular, square and hexagonal nanogrids can be derived by summing up the average $|E/E_0|^4$ from LSPR1 (TM mode) and LSPR2 (TE mode) (Figure 3d) multiplied by their respective coupling coefficients associated with the dimension and symmetry (Section S4, Table S4 and Figure S15). The height ($S_s/S_b$) dependences of the total average $|E/E_0|^4$ for 18, 27 and 36 nm Au/SiO$_2$ nanogrids_$t$200, $s$200 and $h$200 are shown in Figure 4a. The calculated results are found to agree very well with the SERS experiments given in Figure 1d-1f, which reveals the correctness of the models and the validity of the tactful decomposition handling of the polarization direction of incident light and the different nanogrids above.

It has been demonstrated above that the polarization of incident light has a definite effect on both LSPR and SPP effects and thus on SERS effects. Therefore, the polarization dependence of structure - associated SERS effects would also exert an influence on the detection uniformity of a probe. Polarization (polarization angle, $\theta$ is defined as the angle between the polarization of incident light and $x$-direction) dependences of intrinsic SERS effects for $t$200, $s$200 and $h$200, along with their corresponding experimental ones are shown in Figure 4b for comparison.

Theoretical results describe the experiments very well. For both $t200$ and $h200$ nanogrids very weak $\theta$ dependences are observed, which relates closely to the 6-fold symmetry involved in their structures. However, the symmetry with respect to $\theta = 45°$ is seen for $s200$ with the maxima at both $\theta=0°$ and 90°, and the minimum at $\theta=45°$, which is easily understood in terms of its 4-fold symmetry (Figure S15). The structures with a 6-fold symmetry show the far weaker polarization dependences of structural coefficient compared to those with a 4-fold symmetry. The polarization dependences of intrinsic SERS effects are determined by the polarization dependences of structural coefficients which are closely coupled to the symmetry of structures (Table S4 and Figure S15). Therefore, the structures with a higer symmetry show a weaker polarization dependence of intrinsic SERS effects, which would undoubtedly promote the detection uniformity.

So far, very good agreement between theory and experiment for both height and polarization dependences of intrinsic SERS effects verifies the methodology proposed in this study. The intrinsic average and maximum $|E/E_0|^4$ determine the normalized SERS intensity. For the detection of sufficiently high concentration of molecules, the average $|E/E_0|^4$ normally plays a key role because the majority of hot spots decorated with molecules contribute to the SERS intensity. However, for the extremely low concentration of molecules (trace molecules) the hot spots with the maximum $|E/E_0|^4$ which determines the detection limit would play a far more dominant role. Therefore, for each sidewall spacing $L_w$ there are an average $|E/E_0|^4$ and a maximum $|E/E_0|^4$ available in which the former means something for the high concentration of molecules and the latter is important for the extreme low concentration of molecules. Thus, we calculated sidewall spacing $L_w$ dependences of both the average and maximum $|E/E_0|^4$ for triangular, square and hexagonal 36 nm Au/198 nm SiO$_2$ nanogrids with all multiple coupling effects considered, which are presented in Figure 4c and 4d. For all the nanogrids both the average and maximum $|E/E_0|^4$ achieve the maxima at $L_w$ = 302 nm i.e. $\lambda_{\text{SPP Au/air}}$/2 in which the summit of the average $|E/E_0|^4$ for the square nanogrids are the highest. In contrast, the summit of the maximum $|E/E_0|^4$ for the hexagonal nanogrids with a polarization angle of 30° is the highest owing to its maximum SPP coupling effects. Thus, the hexagonal 36 nm Au/198 nm SiO$_2$ nanogrids_$h$188 with the polarization angle of 30° should have the best detection limit theoretically. The total contributions from SPP1 and SPP2 coupling effects are given in Figure 4e and 4f, respectively, and their respective contributions are shown in Figure S16. The maximum contributions of SPP coupling effects for the hexagonal nanogrids at $L_w$ = $\lambda_{\text{SPP Au/air}}$/2 with the polarization angle of 30° are 58.2% and 59.0% for the average and maximum $|E/E_0|^4$, i. e. 139% and 144% those without these effects, respectively, independent of Au roughness but related to plasmonic metals, beyond the conventional recognition. The optimization of structure and dimension can lead to at least one order of magnitude increase in the maximum $|E/E_0|^4$, as shown in Figure 4d, which is expected to improve the detection limit significantly. The quantitative analysis of multiple coupling effects in 3D Au/SiO$_2$ periodic nanogrids is given in Table S5.

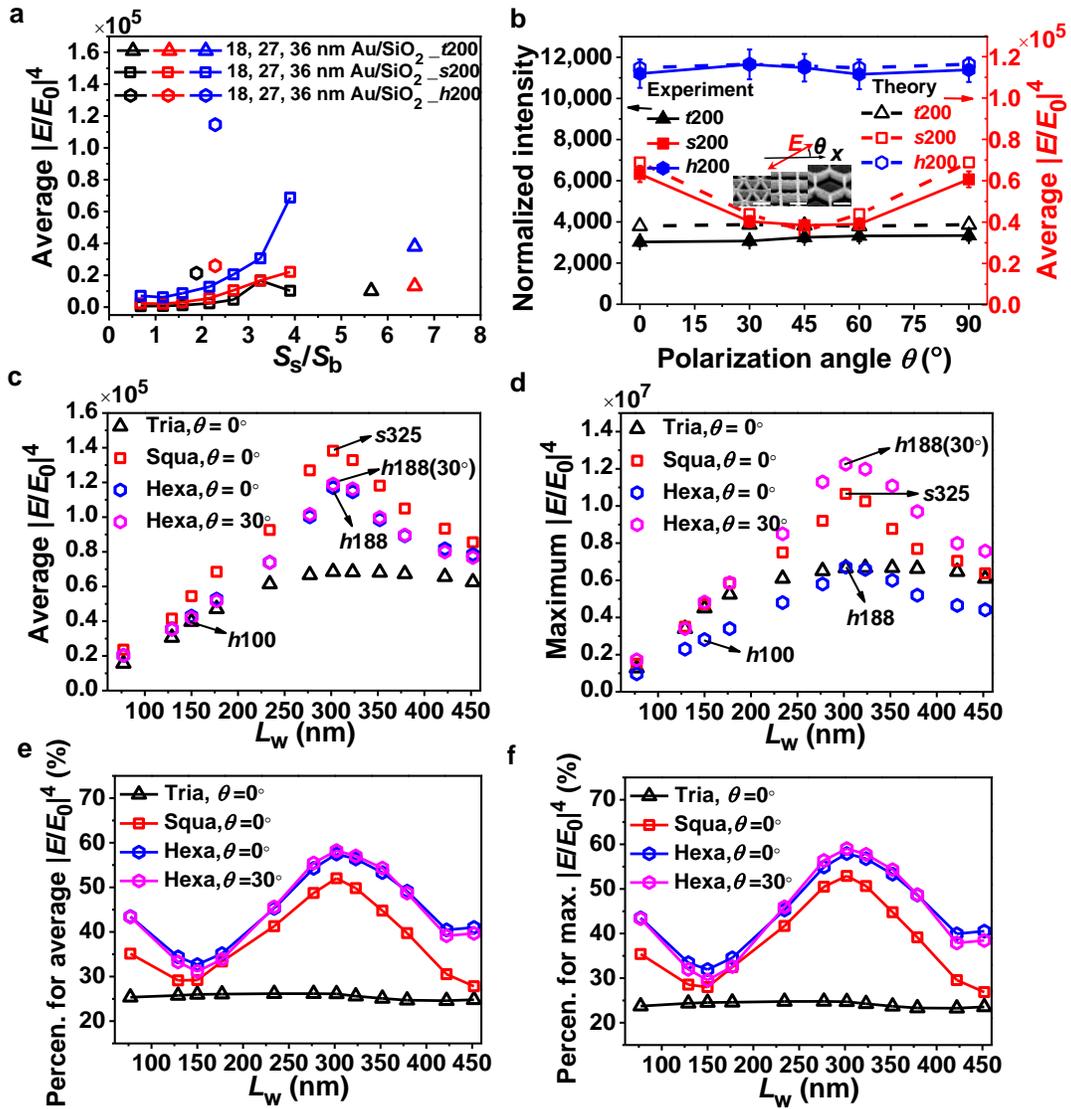

**Figure 4.** Height dependences of the total average $|E/E_0|^4$ for different nanogrids with multiple coupling effects considered, experimental and calculated polarization dependences of SERS effects for different nanogrids with different symmetries, and calculated sidewall spacing $L_w$ dependences of total average and maximum $|E/E_0|^4$ and contributions of SPP coupling excitation effects. (a) Calculated total average $|E/E_0|^4$ versus height ($S_s/S_b$) for triangular, square and hexagonal nanogrids with 200 nm grid length, 18 nm Au/163 nm SiO$_2$, 27 and 36 nm Au/198 nm SiO$_2$, respectively and square 18, 27 and 36 nm Au/SiO$_2$ nanogrids_s200 for the *x*-polarized light with multiple coupling effects considered. (b) Comparison of polarization angle $\theta$ dependences of the normalized experimental Raman intensities and the calculated average $|E/E_0|^4$ for 36 nm Au/198 nm SiO$_2$ nanogrids_t200, s200 and h200. Experiments and calculations agree very well. The polarization angle $\theta$ is defined in the inset. Scale bars: 200 nm. (c) and (d) Sidewall spacing $L_w$ dependences of the calculated average and maximum $|E/E_0|^4$, respectively for triangular,

square and hexagonal 36 nm Au/198 nm SiO$_2$ nanogrids with multiple coupling effects considered. (e) and (f) Sidewall spacing dependences of the total contributions of SPP wave coupling excitation effects to the average and maximum $|E/E_0|^4$, respectively.

Thus far, quantitative dependences of multiple LSPR and SPP effects on geometric dimension, symmetry, polarization and roughness are presented from which a generalized methodology can be developed to accurately design the dimensions of structures with maximum coupling effects. The optimized *s*325 and *h*188 samples based on the square and hexagonal 36 nm Au/198 nm SiO$_2$ nanogrids, respectively, as SERS probes are compared with *s*174, *s*475, *h*100, and *h*274 with their SEM images in Figure S17. Their theoretical average $|E/E_0|^4$ and experimental normalized intensities agree quite well, as shown in Figure 5a. The SERS mapping for the optimized *h*188 gives a low RSD of 5.56% due to its high symmetry (Figure S18), normally much smaller than those of the reported SERS substrates [9,40,41] which is very significant for SERS probes. Theory reveals that *h*188 with a polarization angle of 30° (*h*188 (30°)) shows the maximum $|E/E_0|^4$ in Figure 4d and thus has the best detection limit. As a result, R6G with the concentration of 2.5×10$^{-11}$ M can be detected only by *h*188 (30°) with 40 times improvement of detection limit compared to that of *h*100 albeit with the far higher density of hot spots (Figures 4d, 5a-5e and Figure S17), and a SERS enhancement factor (SERS EF) of 3.4×10$^8$ (Figure 5e and Section S5), which proves theory well, being outstanding among those reported.[9,22,41,42] Surprisingly, the detection limit of R6G changes as *h*100 > *h*188 > *s*325 > *h*188 (30°), which agrees very well with the change of the maximum $|E/E_0|^4$ in Figure 4d. Though *s*325 has the highest average $|E/E_0|^4$ in Figure 4c its detection limit is just 5.0×10$^{-11}$ M and not the best, further demonstrating that the detection limit is determined by the maximum $|E/E_0|^4$ (Figure 4d) instead of the average $|E/E_0|^4$, i. e. the maximum intrinsic EMF enhancement. Therefore, we may design different SERS probes for different application purposes upon requirements of different performances. Here, *h*188 with an intermediate detection limit successfully detected a Hg ion concentration as low as $5.0 \times 10^{-11}$ M (10 ppt), about two orders of magnitude lower than US standard value (10 nM or 2000 ppt) for drinkable water (see Figure 5f and Experimental Section). Therefore, the designability of SERS probes using the methodology proposed here and their practical applications are successfully demonstrated.

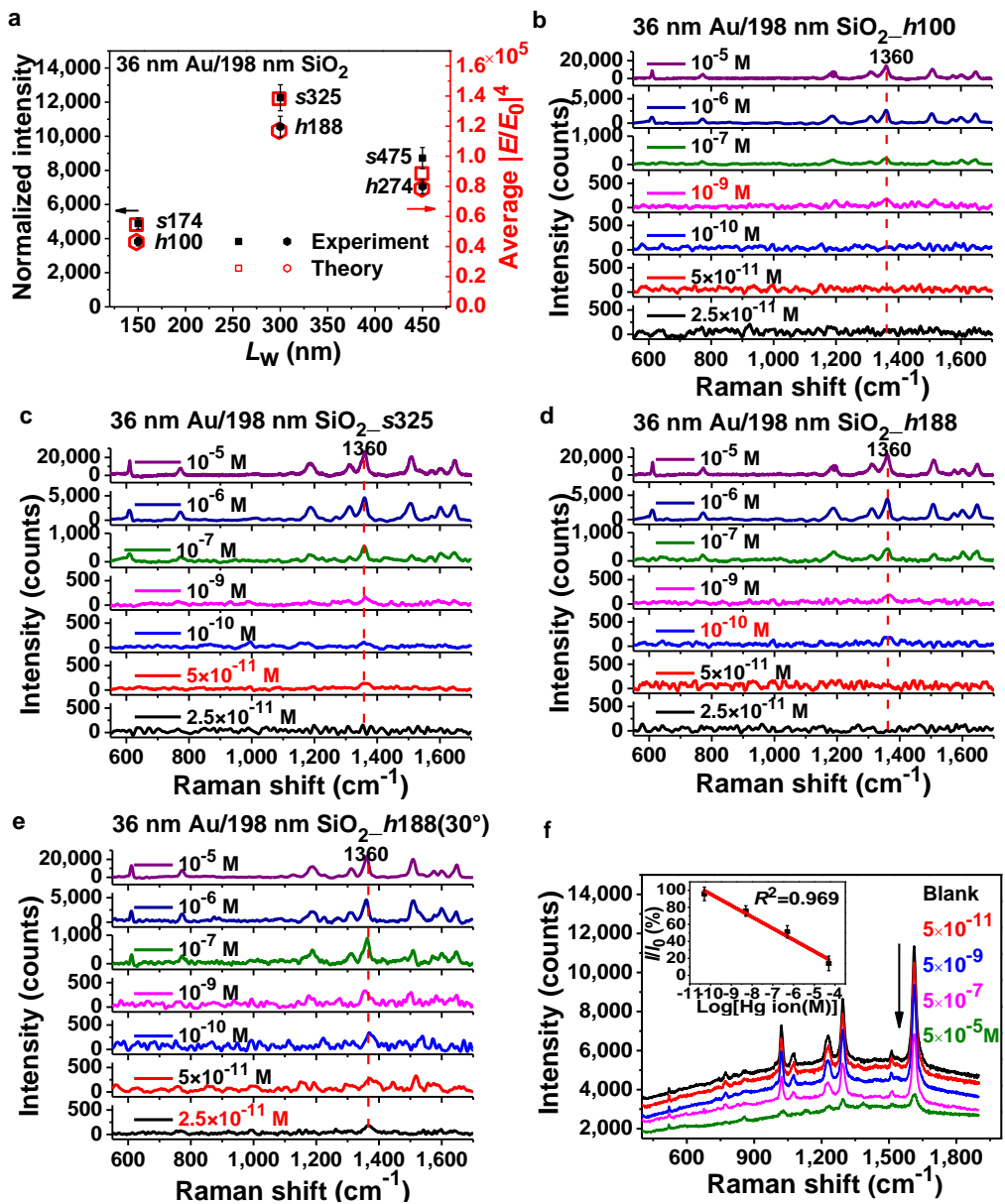

**Figure 5.** Applications of designed SERS probes to detection of trace R6G molecules and Hg ions. (a) Comparison of the normalized Raman intensities of the peak at 1360 cm$^{-1}$ for R6G molecules with concentration of $10^{-5}$ M and calculated average $|E/E_0|^4$ for square nanogrids_$s$174, $s$325 and $s$475, and hexagonal nanogrids_$h$100, $h$188 and $h$274. (b) - (e) Raman spectra of R6G molecules with concentrations from $2.5\times10^{-11}$ to $10^{-5}$ M decorated on 36 nm Au/198 nm SiO$_2$ nanogrids of $h$100, $s$325, $h$188 and $h$188 ($\theta = 30°$), respectively. The detection limits of these SERS probes are concentrations of $10^{-9}$, $5\times10^{-11}$, $10^{-10}$ and $2.5\times10^{-11}$ M, respectively. (f) Detection of heavy metal Hg ions simply by $h$188 probe using 4,4'-Bipyridine (Bpy) molecules with a detection limit as low as $5.0\times10^{-11}$ M (10 ppt), far lower than the value of Hg ion for drinkable water set by U.S. EPA.

**CONCLUSION**

In summary, we develop successfully a generalized method of quantifying formidable multiple coupling effects by deriving dimension, symmetry, Au roughness and polarization dependences of intrinsic EMF enhancements from novel 3D Au/SiO$_2$ periodic hybrid nanogrid models. The optical standing wave effect leads to the EF fluctuating distribution along the height direction of a structure by coupling with LSPR of Au nanoparticles at the sidewalls in which the height with 36% wavelength of incident light is adequate for the achievement of the maximum intrinsic EF enhancement. The SPP waves are excited at both the sidewalls and bottom to produce the FP resonance - like interference effects which can further excite extra LSPR as secondary sources. Both the SPP wave coupling excitation effects and LSPR effects relate to the symmetry of nanostructures due to the polarization dependences in which higher symmetry is preferred to improve the detection limit. For a particular optimized structure the contribution of extra LSPR effects excited by SPP interference waves to the maximum and average intrinsic EF enhancements are even larger than that without the interference of SPP waves, and their contribution ratio is independent of Au roughness, beyond the conventional recognition. The generalized methodology proposed based on the novel findings here can be applied for the exact design of high performance SERS probes by maximizing the contributions of the optical standing wave effects of incident light, SPP interference effects, LSPR effects and their coupling effects to intrinsic EMF enhancements, which is well demonstrated by SERS experiments. The methodology not only enables the accurate dimension design of high performance SERS probe structures but also provide the general design principles qualitatively as follows: 1. Fundamental height requirement for the strongest intrinsic EMF enhancement (A height with 36% wavelength of incident light is adequate for the achievement of the maximum intrinsic EMF enhancement along the height direction, as shown in Figure S10) due to the optical standing wave effect. The higher the height is, the better the intrinsic EMF enhancement is not always; 2. Creation of rough metal surfaces and periodic structures with parallel sidewalls to guarantee the generation of SPP waves to excite extra LSPR further; 3. Adoption of structures with high symmetry for the improvement of detection uniformity; 4. Determination of a particular polarization direction relative to structure to enhance maximum intrinsic EMF for the optimization of detection limit; 5. Increase of metal surface roughness and narrowing of nanogaps as possible to increase intrinsic EMF, which is independent of four items above; 6. Applicable to the combinations of any plasmonic metal / dielectric hybrids which determine the wavelengths of SPP excited. Hexagonal 3D Au/SiO$_2$ periodic nanogrids _$h$188 (30°) designed using the methodology can detect the limit concentration of $2.5 \times 10^{-11}$ M for R6G with 40 times improvement compared to $h$100 with far higher density of hot spots, $3.4 \times 10^8$ SERS EF and RSD 5.56% detection uniformity, and the similar probe designed can successfully detect trace Hg ions in water with $5.0 \times 10^{-11}$ M concentration. Therefore, this study not only provides novel 3D Au/SiO$_2$ periodic nanogrids as SERS probes with high performance, but more importantly addresses the formidable issues of quantifying multiple coupling effects involved in SERS. The generalized methodology proposed here enables the exact design of 3D SERS probe structures with high performance by combining state of the art top down nanofabrication (even far cheaper nanoimprinting or 3D printing) and bottom up technologies.

**Experimental Section**

**Design and Fabrication of Nanogrids.** Three dimensional (3D) Au/SiO$_2$ periodic nanogrids with various geometries and dimensions were designed and fabricated. Hydrogen silsesquioxane HSQ (XR-1541-006, Dow Corning, USA) was firstly spin-coated on silicon (100) substrates with the thicknesses of 34, 58, 79, 108, 134, 163 and 198 nm for height control of nanogrids. Patterning was realized by using Electron - beam lithography (EBL, Vistec EBPG 5000 plus ES, Raith Company, Germany) with an accelerating voltage of 100 kV and a beam current of 2 nA, followed by the development. The typical width of SiO$_2$ sidewalls for all nanogrids was controlled to be around 9 nm. Deposition of Au with the thicknesses of 1, 9, 18, 27 and 36 nm, along with 3 nm thick Cr adhesion layer was performed on an electron-beam evaporator (OHMIKER-50B, Cello-Tech Company, Taiwan, China). As a reference sample, lift-off was conducted by immersing the samples in a 1:5 buffered hydrofluoric acid (HF) solution (7:1 of 40% NH$_4$F and 49% HF) at room temperature with ultrasonic agitation for 4 min. The fabricated structures were observed using a scanning electron microscope (NOVA NanoSEM 430, FEI Company, USA). Atomic force microscopy (AFM) images were acquired on a Veeco Dimension 3100 microscope (Veeco Digital Instruments, US). The schematic of sample fabrication is shown in Figure S1.

**SERS Measurements.** All the nanogrids with different symmetries and geometric dimensions were firstly immersed into Rhodamine-6G (R6G) aqueous solution with the concentrations ranging from $2.5\times10^{-11}$ to $10^{-5}$ M for 12 h, and then dried naturally in air as SERS substrates. SERS measurements were performed on a Raman spectroscopy (Renishaw inVia, Renishaw company, UK) with a 50× objective (numerical aperture, NA=0.75), a laser of 632.8 nm with a power of 0.5 mW, the *x*-polarization and an integration time of 10s.

**FDTD Calculations and Optical Measurements.** Finite-difference time-domain (FDTD) method was used to calculate the spatial distributions of the electromagnetic fields. For simplicity, we used the rough $Au/SiO_2$ models with periodically arranged hemisphere- or semiellipsoid - like Au nanoparticles on the sidewalls of $SiO_2$ nanogrids to model the real $Au/SiO_2$ nanogrids in which the sizes of Au particles were basically derived from SEM observations. Periodic boundary conditions for the *xz* and *yz* planes were applied to simulate an infinite array of periodic nanogrids or nanowalls. Perfectly matched layer (PML) boundary conditions were used in the *z*-direction. The mesh size used in the simulation region was 2 nm for the calculations of SPP effects and 0.5 nm for the calculations of LSPR effects. The optical constants of exposed HSQ were determined using a spectroscopic ellipsometer (SE 850 DUV, Sentech Company, Germany). The infrared spectra were recorded on a Fourier transform infrared spectrometer (Nicolet iN10, Thermo Fisher Company, USA).

**Detection of Hg Ions.** The samples were first immersed into 4,4'-Bipyridine (Bpy) absolute ethanol solution with a concentration of $10^{-5}$ M for 4 h, and then dried naturally in air as SERS probes for Hg ions detection. 35 $\mu$L of Hg ion solutions with different concentrations of 5.0×10$^{-12}$ (1 ppt), 5.0×10$^{-11}$ (10 ppt), 5.0×10$^{-9}$, 5.0×10$^{-7}$ and 5.0×10$^{-5}$ M was dropped onto the SERS probes, respectively, then kept for 10 min, and finally dried in air. Likewise, 35 $\mu$L of deionized water was prepared with the same procedure as SERS probes for the blank control group. The SERS measurements were performed using a 632.8 nm laser with a power of 1mW and the *x*-polarization on a Renishaw inVia Raman microscope equipped with a 20× objective (NA=0.4) and an integration time of 10 s. For each sample, measurements on at least five different positions were taken.

ASSOCIATED CONTENT

**Supporting Information**.

All the abbreviations; fabrication schematics and geometries of 3D Au/SiO$_2$ periodic nanogrids; characterization analysis for exposed HSQ; SEM images of SiO$_2$, Au/SiO$_2$ nanogrids and reference samples without sidewalls; geometry parameters of Au/SiO$_2$ nanogrids derived from statistical analysis of SEM images; experimental Raman intensities of Au/SiO$_2$ nanogrids with different Au thicknesses and SiO$_2$ heights; calculations of SPP wavelength and SPP waves interference effects; spatial distributions of the electromagnetic field components on the sidewall surfaces calculated by FDTD; the relationships between SPP1 wave interference effects with sidewall length and polarization angle $\alpha$; optical standing wave effect of the incident light; calculated SiO$_2$ height dependences of the average squared electric field intensities and interference intensities of SPP2 wave; geometry parameters of hemispheres / semiellipsoids Au nanoparticles; calculated spatial distributions of the electric field intensities of roughness nanowalls calculated by FDTD; the relationships between EF enhancements of LSPR1 and LSPR2 with polarization angle $\alpha$; coupling coefficients of different nanogrids; sidewall spacing dependences of the contributions of SPP1 and SPP2 wave coupling excitation effects respectively; comparison of theoretical results of multiple effects of Au/SiO$_2$ nanogrids; SEM images of square and hexagonal 36 nm Au/198 nm SiO$_2$ nanogrids with different sidewall spacings; Raman intensity mapping; calculation of SERS enhancement factor (PDF)

## AUTHOR INFORMATION

**Author Contributions**

Y.T., W.C. and P.C. conceived the idea; P.C., L.Y. and Y.T. fabricated the SERS probes; Y.T., F.D. and H.W. performed the simulations; Y.T., X.Z., Y.G. and L.S. performed the measurements; W.C., P.C., Y.T., H.W., A.F. and L.S. supervised the analysis, experiments, and edited the manuscript.

**Notes**

The authors declare no competing financial interest.

## ACKNOWLEDGMENT

This work was financially supported by the project "Exploration of High Performance Nanosensors for Rapid Detection and Application Demonstration" of The National Key Research and Development Program of China "Fundamental Research on Nano Sensing Materials and High Performance Sensors Focused on Pollutants Detection" under grant No. 2017YFA0207104, the Strategic Priority Research Program of the Chinese Academy of Sciences under grant No. XDA09040101, National Natural Science Foundation of China under grant No. Y6061111JJ, Youth Innovation Promotion Association CAS under grant No. 2015030, and CAS Key Technology Talent Program under grant Nos. Y8482911ZX and Y7602921ZX.

Supporting Information:

Design of Novel 3D SERS Probes with Drastically Improved Detection Limit by Maximizing SPP – Based Multiple Coupling Effects

*Yi Tian,*[†,‡] *Hanfu Wang,*[†] *Lanqin Yan,*[†] *Xianfeng Zhang,*[†] *Attia Falak,*[†,‡] *Yanjun Guo,*[†] *Peipei Chen,\**[,†] *Fengliang Dong,\**[,†] *Lianfeng Sun,\**[,†,‡] *and Weiguo Chu\**[,†,‡]

[†] CAS Key Laboratory for Nanosystems and Hierachical Fabrication, Nanofabrication Laboratory, CAS Center for Excellence in Nanoscience, National Center for Nanoscience and Technology, Beijing 100190, P. R. China

[‡] University of Chinese Academy of Sciences, Beijing 100049, P. R. China

**Corresponding Author**

* E-mail: wgchu@nanoctr.cn.   * E-mail: chenpp@nanoctr.cn.   * E-mail: dongfl@nanoctr.cn.

* E-mail: slf@nanoctr.cn.

**Table S1.** All the abbreviations and the corresponding indications.

| Abbreviations | Indications |
|---|---|
| 18, 27, 36 nm Au/198 nm SiO$_2$_$h$200 (_$s$200, _$t$200) | 18, 27 and 36 nm thick Au evaporated on 198 nm high SiO$_2$ hexagonal (square, triangular) nanogrids with a 200 nm grid length, respectively |
| $h$100, $h$200, $h$188, $h$188 (30°) and $h$274 | Hexagonal nanogrids with a grid length of 100, 200, 188, 188 (with a polarization angle of 30°) and 274 nm, respectively, for all different Au thicknesses and SiO$_2$ heights |
| $s$174, $s$200, $s$325 and $s$475 | Square nanogrids with a grid length of 174, 200, 325 and 475 nm, respectively, for all different Au thicknesses and SiO$_2$ heights |
| $t$200 | Triangular nanogrids with a grid length of 200 nm, for all different Au thicknesses and SiO$_2$ heights |
| 18, 27 and 36 nm Au / SiO$_2$ nanowalls_$D$200 | 18, 27 and 36 nm thick Au evaporated on SiO$_2$ hexagonal (square, triangular) nanowalls with a wall center distance of 200 nm and different heights, respectively |
| 36 nm Au/198 nm SiO$_2$ nanogrids | Triangular, square and hexagonal 36 nm thick Au on 198 nm high SiO$_2$ nanogrids with different grid lengths |

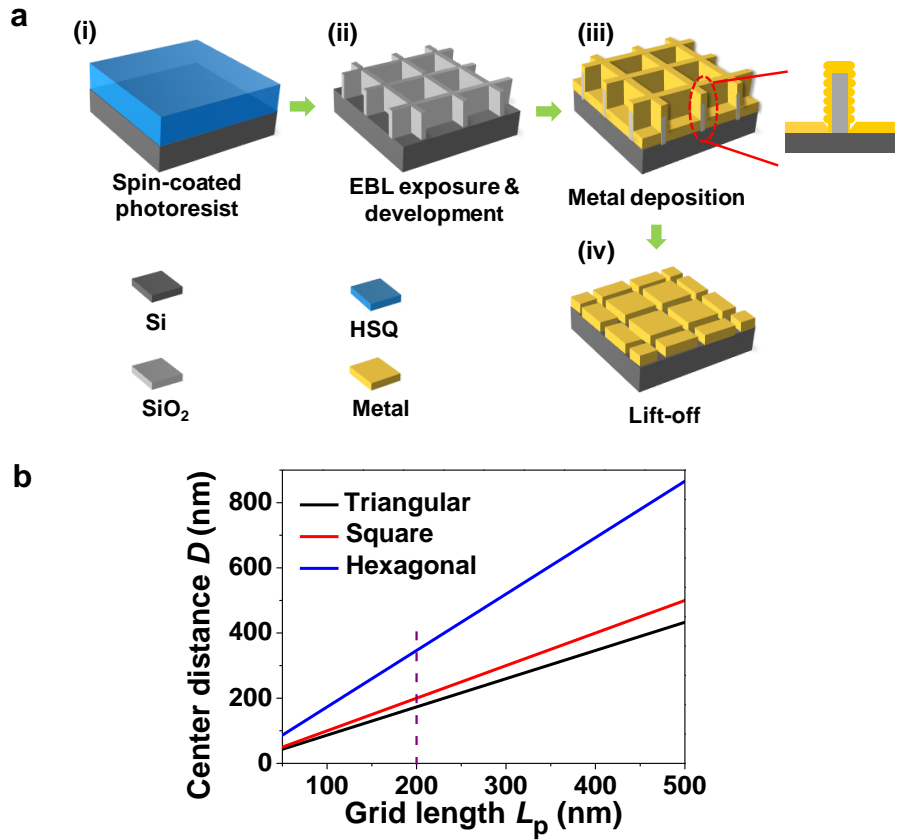

**Figure S1. Fabrication schematics of 3D Au/SiO$_2$ periodic nanogrids, their corresponding nanostructures without sidewalls as the references and the relationships between grid length $L_p$ and center distance $D$. a**, Firstly, HSQ resist was spin-coated on Si substrates (i), followed by EBL and development (ii) according to designs. The development of exposed HSQ would convert the HSQ to SiO$_2$. Afterwards, ultrathin Cr adhesion layers and Au films with various thicknesses were deposited using electron beam evaporation (iii). Finally, the corresponding nanostructures without sidewalls were prepared as the references for comparison by using a lift-off process (iv). **b**, The relationships between grid length $L_p$ and sidewall center distance $D$ of triangular, square and hexagonal Au/SiO$_2$ nanogrids.

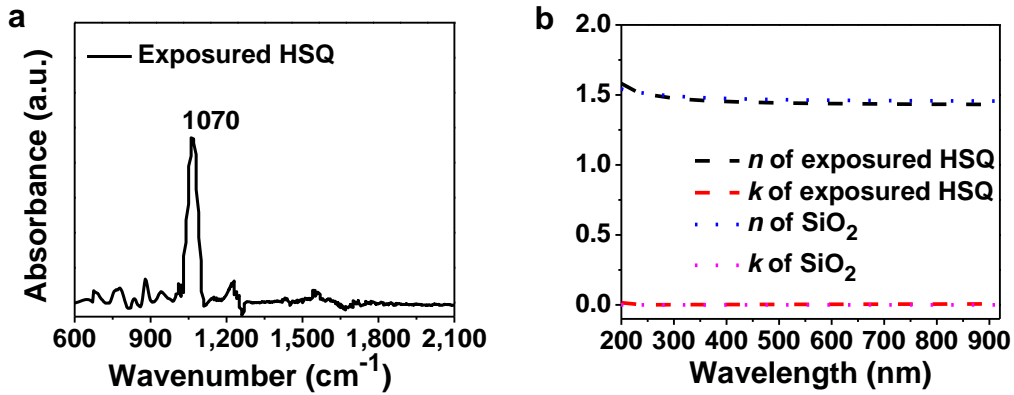

**Figure S2. Fourier transform infrared spectrum of large-area exposed HSQ and comparison of its optical constants and those of SiO$_2$. a**, Fourier transform infrared (FTIR) spectrum of exposed HSQ, where 1070 cm$^{-1}$ peak is a typical absorption peak of Si–O–Si network structures, characterized by the presence of SiO$_2$.[1,2] **b**, Comparison of optical constants of exposed HSQ derived from the measurements by spectroscopic ellipsometry and typical SiO$_2$.

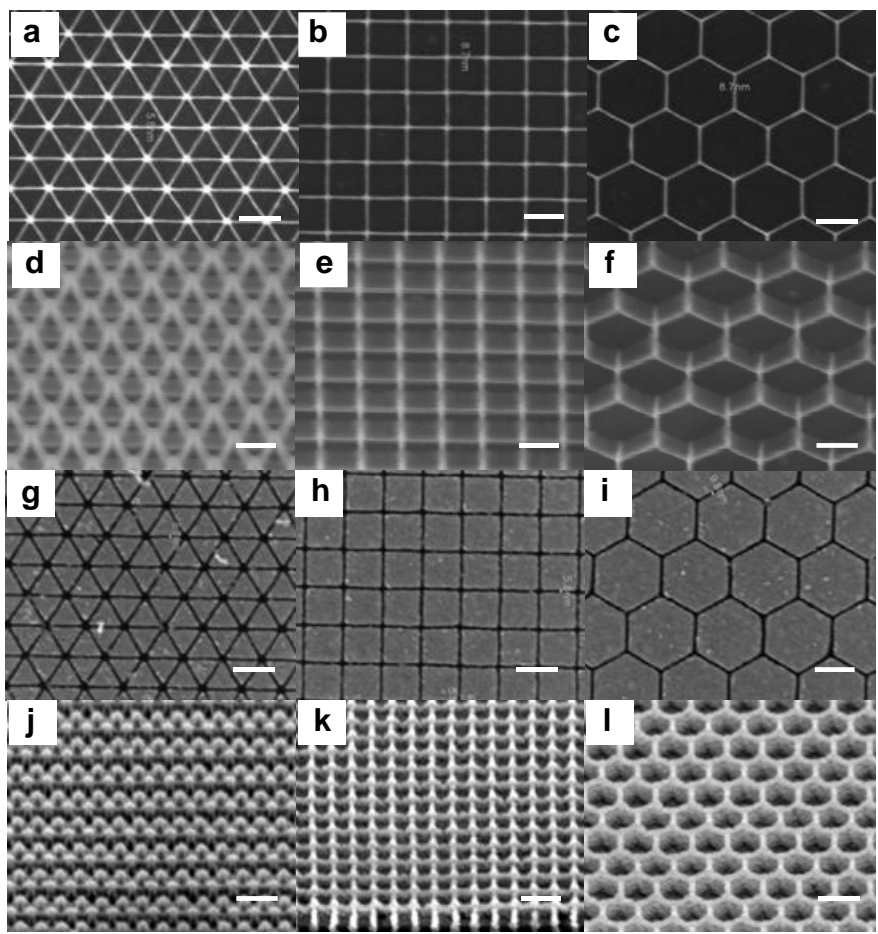

**Figure S3. SEM images of triangular, square and hexagonal periodic SiO$_2$, Au/SiO$_2$ nanogrids and corresponding reference samples without sidewalls. a-c**, SEM images of triangular, square and hexagonal SiO$_2$ nanogrids with 198 nm heights and 200 nm grid length. **d-f**, Corresponding tilt SEM images in **a-c**. **g-i**, SEM images of triangular, square and hexagonal 36 nm thick Au periodic nanostructures without sidewalls with 200 nm side length prepared by lift-off as the reference samples. **j-l**, Tilt SEM images of triangular, square and hexagonal 36 nm Au/198 nm SiO$_2$ nanogrids with 100 nm grid length. Scale bars: 200 nm.

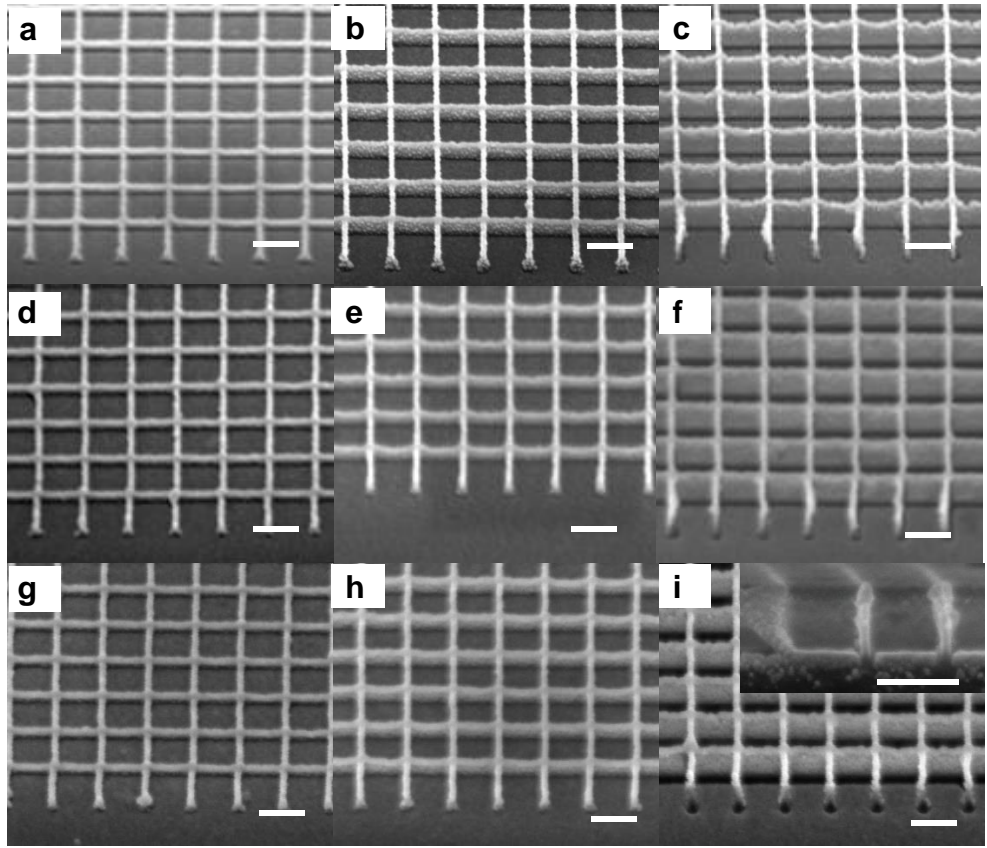

**Figure S4. Tilt SEM images of square Au/SiO$_2$ nanogrids with 200 nm grid length, different Au thicknesses and SiO$_2$ heights. a-c**, 18 nm Au/SiO$_2$ nanogrids with 58, 108 and 198 nm height, respectively. **d-f**, 27 nm Au/SiO$_2$ nanogrids with 58, 108 and 198 nm height, respectively. **g-i**, 36 nm Au/SiO$_2$ nanogrids with 58, 108 and 198 nm height, respectively. The inset in **i** shows SEM image of sidewalls of the 36 nm Au/198 nm SiO$_2$ nanogrids. Noting that with increased height and Au thickness the roughness tends to increase initially and decrease then. Scale bars: 200 nm.

**S1. Geometry parameters of Au/SiO₂ nanogrids derived from statistical analysis of SEM images**

The average width $W_0$ of SiO$_2$ sidewalls, the width $W_1$ and height $H_1$ of square Au/SiO$_2$ hybrid sidewalls with 200 nm grid length which are defined in Figure S5a were statistically extracted from SEM images (Figure S4). SiO$_2$ height $H_0$ dependences of these parameters and the cross sectional Au areas of sidewalls for 18, 27 and 36 nm thick Au without considering the roughness are constructed in Figure S5. The average width $W_0$ of bare SiO$_2$ sidewalls is found to be around 9 nm, and the width $W_1$ of hybrid sidewalls is between 15 and 25 nm for 18, 27 and 36 nm Au thickness, and tends to decrease with the increased $H_0$ for an Au thickness, as shown in Figure S5b. The average height $H_1$ of Au/SiO$_2$ hybrid nanogrid sidewalls and the height $H_0$ of bare SiO$_2$ nanogrid sidewalls are almost same (Figure S5c). Based on the geometrical relationship shown in Figure S5a, SiO$_2$ height $H_0$ dependences of the cross sectional Au area of sidewalls can be described as $S = H_1 (W_1 - W_0) \approx H_0 (W_1 - W_0)$, which are well fitted by the second order polynomial $S(H_0) = A + BH_0 + CH_0^2$ (Figure S5d) with the corresponding $A$, $B$ and $C$ values outlined in Table S2.

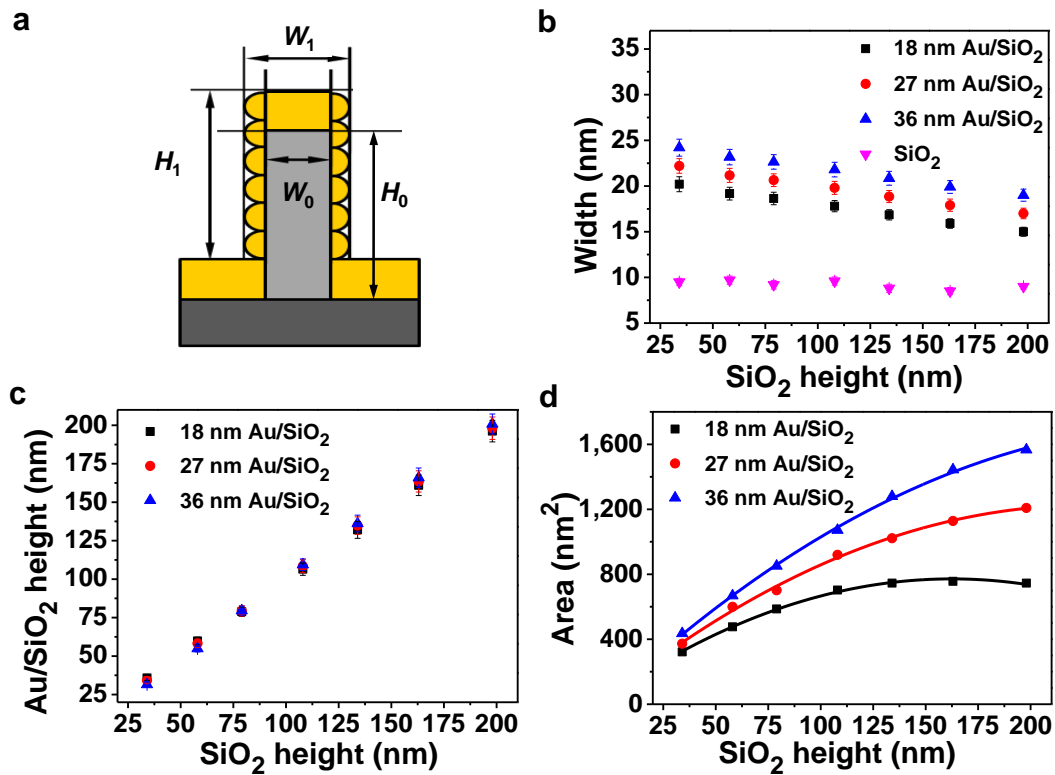

**Figure S5. Geometry parameters of square 18, 27 and 36 nm Au/SiO$_2$ nanogrids with 200 nm grid length and various SiO$_2$ heights. a**, Geometry parameters of nanogrids defined. **b**, Changes of average sidewall width $W_1$ of square Au/SiO$_2$ nanogrids and sidewall width $W_0$ of bare SiO$_2$ nanogrids with height. **c**, The relationships between the average height $H_1$ of Au/SiO$_2$ hybrid nanogrids and the height $H_0$ of bare SiO$_2$ nanogrids. **d**, The cross sectional Au areas of sidewalls versus the height of bare SiO$_2$ nanogrids from which the densities of Au nanoparticles can be derived for FDTD calculations.

**Table S2.** Coefficients of second order polynomial employed to fit the relationship between the cross section areas of Au sidewalls and height presented in Figure S5d.

| Au thickness (nm) | *A* | *B* | *C* |
|---|---|---|---|
| 18 | 58.73 | 8.77 | -0.027 |
| 27 | 57.13 | 10.26 | -0.0225 |
| 36 | 46.73 | 11.97 | -0.0214 |

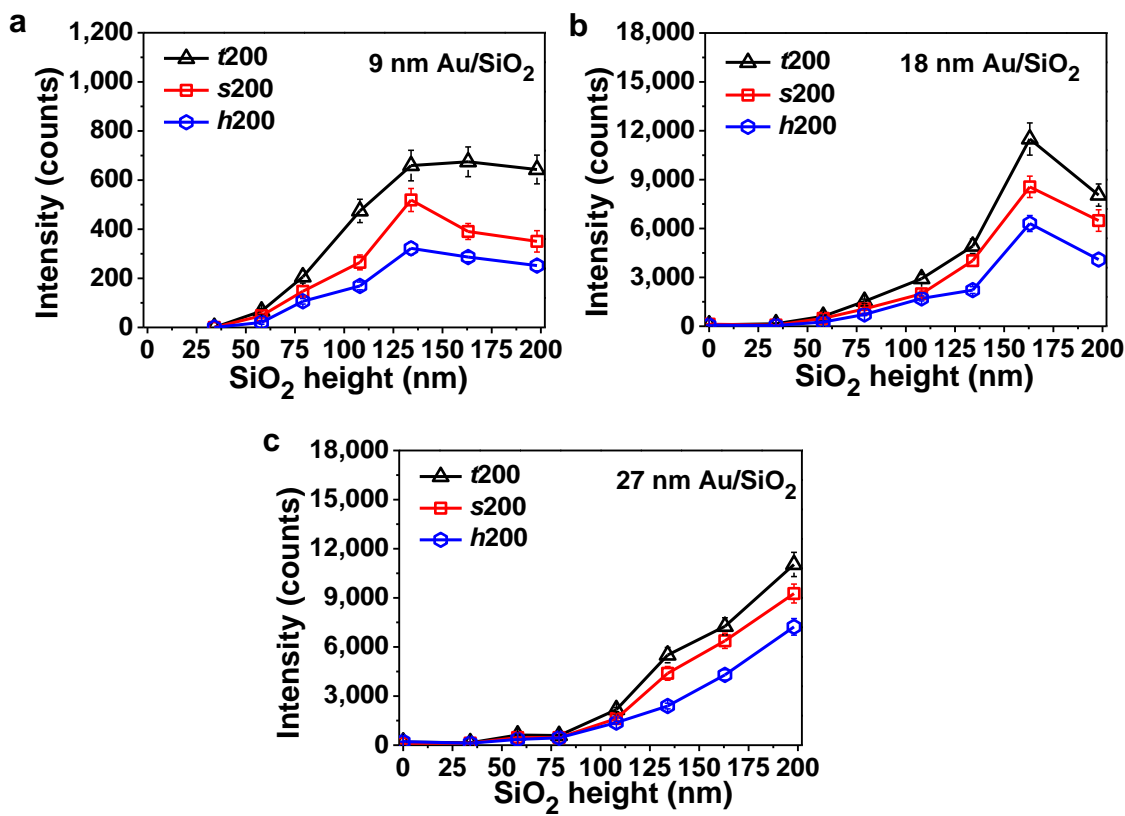

**Figure S6. Changes of experimental Raman intensity of the peak positioned at 1360 cm$^{-1}$ of the R6G - decorated triangular, square and hexagonal Au/SiO$_2$ nanogrids with 200 nm grid length with SiO$_2$ nanogrid height.** The heights are measured to be 34, 58, 79, 108, 134, 163 and 198 nm, respectively, which correspond to the nominal heights of 35, 60, 80, 110, 135, 165 and 200 nm, respectively. **a**, 9 nm thick Au. **b,** 18 nm thick Au. **c**, 27 nm thick Au.

## S2. Calculations of SPP wavelength ($\lambda_{SPP}$) and SPP waves interference effects

It is well recognized that SPP can be excited by an incident light for a rough surface because the conditions of wavevector match for the incident light and SPP can be easily satisfied in the near field region owing to the random reflection in many directions.[3,4] As shown in Figure S7a, SPP1 for the model can be excited by a polarized light (near-field) with a component of its electric field ($E_y$) perpendicular to the surface of sidewalls (in the xz-plane) or ($E_x$) parallel to the propagation direction of SPP.[4] SPP2 can be excited by a polarized light with a component of its electric field ($E_x$) perpendicular to the nanowalls (along the y-direction), i.e. TM mode.

SPP on a metal film/dielectric interface propagate in the plane with the magnitude of wave vector [5]

$$k_{SPP} = \frac{\omega}{c}\left(\frac{\varepsilon_m \varepsilon_d}{\varepsilon_m + \varepsilon_d}\right)^{1/2} \quad (1)$$

where $\omega$ and $c$ are the frequency and speed of excitation light in free space, respectively, and $\varepsilon_m$ and $\varepsilon_d$ are the dielectric constants of metal and dielectric material (Au/SiO$_2$ and Au/air here), respectively. Then, the SPP wavelength, $\lambda_{SPP}$, can be obtained as

$$\lambda_{SPP} = \lambda_{ex}\left(\frac{\varepsilon_m + \varepsilon_d}{\varepsilon_m \varepsilon_d}\right)^{1/2} \quad (2)$$

where $\lambda_{ex}$ is the wavelength of excitation source in free space. Accordingly, for $\lambda_{ex}$ =632.8 nm, $\lambda_{SPP\ Au/air}$ for Au/air interface is derived to be 603 nm with $\varepsilon_m$= −10.88 [6] and $\varepsilon_d$=1, and $\lambda_{SPP\ Au/SiO_2}$ for Au/SiO$_2$ interface is derived to be 389 nm with $\varepsilon_m$= −10.88 and $\varepsilon_d$=2.13.

Both SPP1 and SPP2 waves can propagate not only in the positive *x* direction but also in the opposite direction (Figure S7a). The interference of two SPP waves with the opposite directions is not considered because there is no fixed phase difference between them. When the excited SPP wave propagates to encounter a sidewall, the partial reflection, transmission and scattering all would occur. Here, we reasonably consider only the reflected SPP wave because we aim to study the interference of SPP wave confined in a single cavity. Once reflected, the SPP wave propagates again to encounter the sidewall to be partially reflected again. Thus, the multiple reflections of SPP wave take place in a single cavity to create a Fabry - Perot (FP) resonance - like interference with an interference intensity of $I_{\text{SPP-interference}}$. The electric field intensity of SPP wave interference is

$$E_{\text{SPP-interference}}(x) = E_{\text{SPP}}(x) + E_{\text{SPPr1}}(x) + E_{\text{SPPr2}}(x) + E_{\text{SPPr3}}(x) + E_{\text{SPPr4}}(x) + \text{L}$$

$$= A_{\text{SPP}} \left[ \left(\sqrt{R}\right)^n e^{i\left((-1)^n k_{\text{SPP}} x - \omega t + n k_{\text{SPP}} L\right)} \bigg|_{n=0,2,4L} + \left(\sqrt{R}\right)^n e^{i\left((-1)^n k_{\text{SPP}} x - \omega t + \pi + (n+1) k_{\text{SPP}} L\right)} \bigg|_{n=1,3,5L} \right] \quad (3)$$

$$= A_{\text{SPP}} e^{i(-\omega t)} \left[ \left(\sqrt{R}\right)^n e^{i\left((-1)^n k_{\text{SPP}} x + n k_{\text{SPP}} L\right)} \bigg|_{n=0,2,4L} - \left(\sqrt{R}\right)^n e^{i\left((-1)^n k_{\text{SPP}} x + (n+1) k_{\text{SPP}} L\right)} \bigg|_{n=1,3,5L} \right]$$

So, the interference intensity of SPP wave is

$$I_{\text{SPP-interference}}(x) = |E_{\text{SPP-interference}}(x)|^2 = E_{\text{SPP-interference}}(x) E_{\text{SPP-interference}}^*(x)$$

$$= A_{\text{SPP}}^2 \left[ \left(\sqrt{R}\right)^n e^{i\left((-1)^n k_{\text{SPP}} x + n k_{\text{SPP}} L\right)} \bigg|_{n=0,2,4L} - \left(\sqrt{R}\right)^n e^{i\left((-1)^n k_{\text{SPP}} x + (n+1) k_{\text{SPP}} L\right)} \bigg|_{n=1,3,5L} \right] \quad (4)$$

$$\left[ \left(\sqrt{R}\right)^n e^{i\left((-1)^n k_{\text{SPP}} x + n k_{\text{SPP}} L\right)} \bigg|_{n=0,2,4L} - \left(\sqrt{R}\right)^n e^{i\left((-1)^n k_{\text{SPP}} x + (n+1) k_{\text{SPP}} L\right)} \bigg|_{n=1,3,5L} \right]^*$$

where $A_{\text{SPP}}$ is the amplitude of SPP wave, *R* is the reflectance of SPP wave reflected by the opposite sidewall of cavity and *L* is the cavity length.

One should bear in mind that the interference occurring at the hot spots would influence the intensity of SERS. Therefore, we consider the SPP1 wave interference effects at the surface (in the $y=0$ plane) of nanowalls along the $x$-direction ($x$-nanowall) and the SPP2 wave interference effects at the surface of $y$-nanowalls (in the plane with $x=0$) and $x$-nanowalls (in the plane with $y=0$), which can further excite LSPR at the hot spots of the square nanogrid sidewalls besides the incident light (Figure S7a).

For the SPP1 wave interference effect in the plane with $y=0$, we need to average the interference intensity of SPP1 wave from $x=0$ to $x=L=L_s$ (Here, the cavity length equals the sidewall length $L_s$.) as follows

$$\bar{I}_{\text{SPP1-interference}} = \frac{1}{L_s} \int_0^{L_s} I_{\text{SPP1-interference}}(x) \tag{5}$$

For the SPP2 wave interference effect in the plane with $x=0$, the interference intensity of SPP2 is $I_{\text{SPP2-interference}}(0)$. For the SPP2 wave interference effect in the plane with $y=0$, we also need to average the interference intensity of SPP2 wave from $x = 0$ to $x = L = L_w$ (Here, the cavity length equals the sidewall spacing $L_w$.) as follows

$$\bar{I}_{\text{SPP2-interference}} = \frac{1}{L_w} \int_0^{L_w} I_{\text{SPP2-interference}}(x) \tag{6}$$

Upon calculating, $R=0.29$ and $0.13$ were taken from the FDTD calculations on the model with 7 nm thickness of Au on each side of an Au/SiO$_2$ sidewall for the SPP wavelengths of 603 and 389 nm, respectively. We calculated the penetration depth of about 27 nm in Au for Au/SiO$_2$ interface. Based on the exponential decay of intensity, the intensity of SPP1 wave at the Au/SiO$_2$ interface is figured out to be three fifths that of SPP1 at the Au/air interface with the SPP1 wave at the Au/SiO$_2$ interface travelling across 7 nm thick Au considered. Then, we calculated the ratio between the SPP1 wave interference intensity and the SPP1 wave intensity in the $y=0$ plane against sidewall length, and that for SPP2 in the $x=0$ and $y=0$ planes against $L_w/\lambda_{SPP2}$ by only considering the first seven polynomials of SPP wave interference items, as shown in Figure S7b - d. Clearly, the SPP wave interference is found to be similar to Fabry-Perot resonance, and the intensity is a periodic function of the cavity length.

The derivation of the SPP1 wave interference effects is as follows. The ratios of the averaged fourth power of electric field intensities [7,8] (proportional to SERS intensities) at one sidewall of the rough triangular, square and hexagonal 36 nm Au/198 nm SiO$_2$ nanogrids to those of nanowalls, i. e. $(1+I_{SPP1-interference})/(1+I_{SPP1})$ (The intensity of the primary incident light is set to be 1 for all calculations, and $I_{SPP1}$ and $I_{SPP1-interference}$ are SPP1 wave and its interference intensities, respectively), against sidewall lengths $L_s$ can be derived for $\alpha = 0$ and $\pi/2$ ($\alpha$, defined as the angle between the nanowalls and the polarization direction of light) (Figure S8, Supporting Information), as shown in Figure S9a of Supporting Information. Combining $I_{SPP1-interference}/I_{SPP1}$ in Figure S7b of Supporting Information, we derived the sidewall length $L_s$ dependences of $I_{SPP1-interference}$ for $\alpha = 0$ and $\pi/2$ to be shown in Figure 2b and S9b of Supporting Information, respectively.

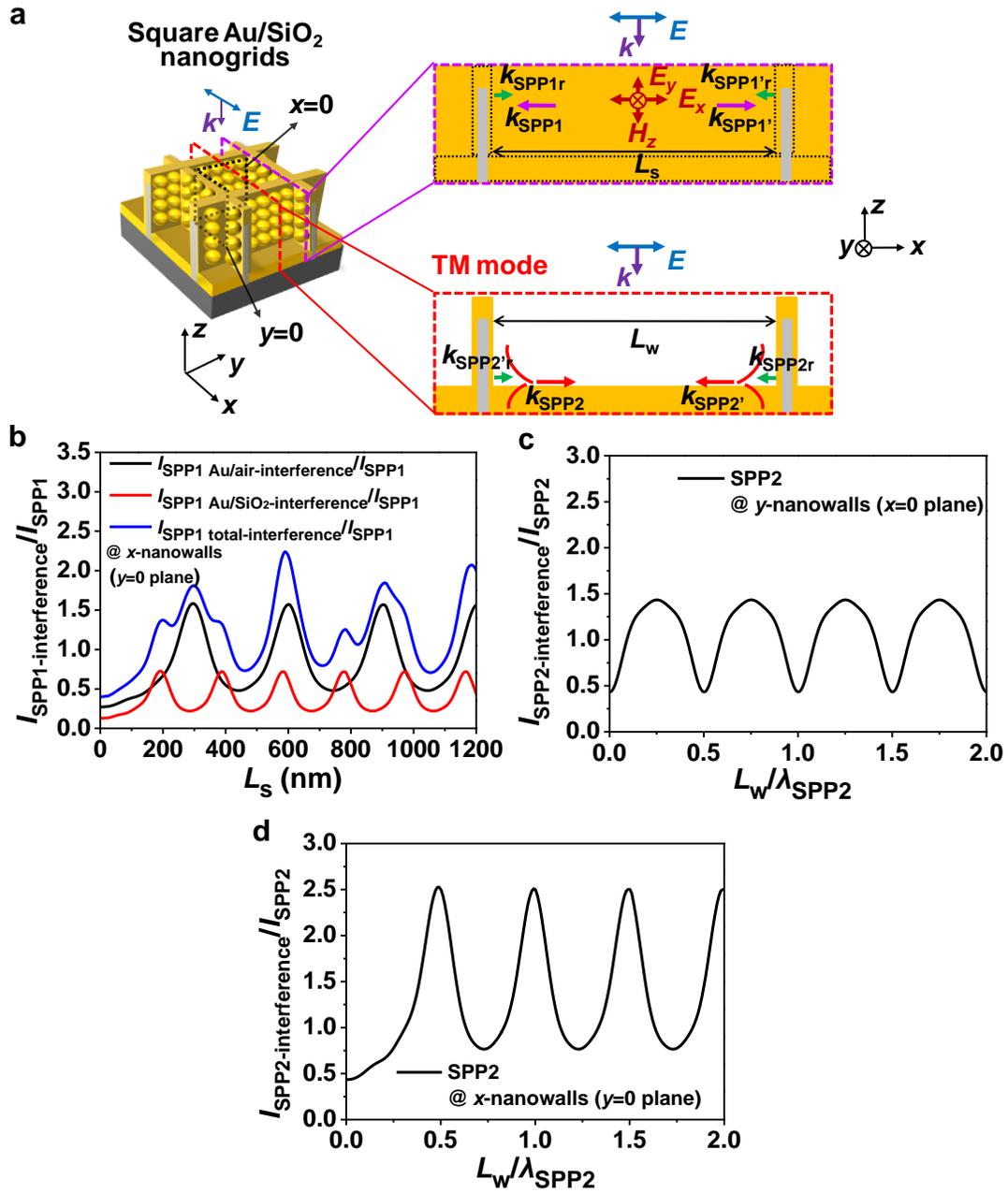

**Figure S7. Mathematical analysis of SPP wave interference effects. a**, Schematics of SPP1 and SPP2 waves interference effects. **b,** Theoretical ratio of the SPP1 wave interference intensity to the SPP1 wave intensity versus sidewall length $L_s$ on the sidewall surface with $y=0$. **c** and **d**, The ratios of the SPP2 wave interference intensity to the SPP2 wave intensity versus $L_w/\lambda_{SPP2}$ on the sidewall surfaces with $x=0$ and $y=0$, respectively, in which $L_w$ is the sidewall spacing.

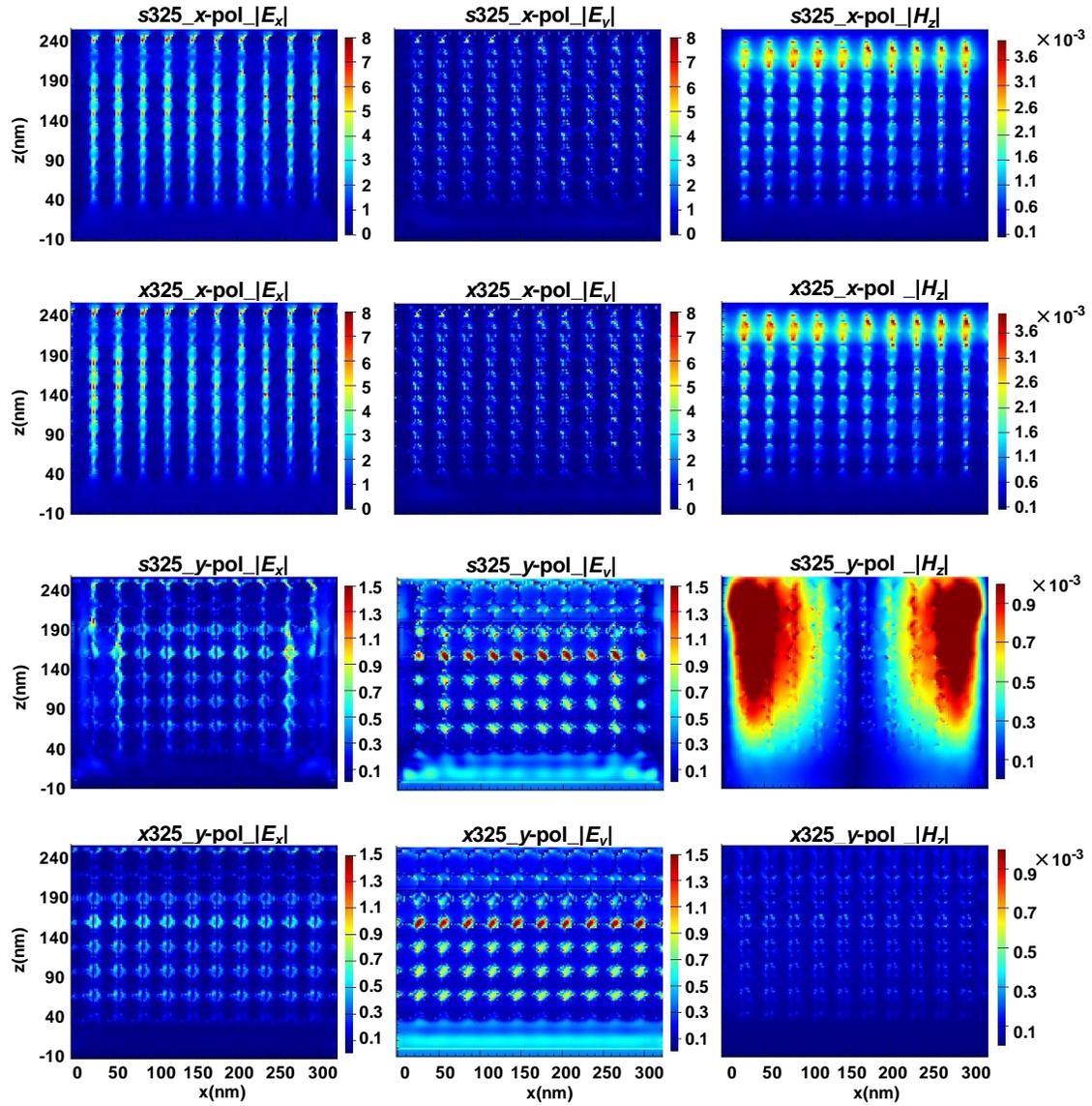

**Figure S8. Calculated spatial distributions of the electromagnetic field components on the sidewall surfaces.** Calculated spatial distributions of $|E_x|$, $|E_y|$ and $|H_z|$ on sidewall surfaces (parallel to the *xz* plane) of rough square 36 nm Au/198 nm $SiO_2$ nanogrids_*s*325 and nanowalls along *x*-direction _*x*325 with 325 nm sidewall center distance *D* (i.e. 302 nm sidewall length with 23 nm sidewall width) for *x* and *y*-polarized light (i.e. for the polarization angles $\alpha = 0$ and $\pi/2$), respectively, by FDTD calculations. It is clear that SPP1 wave can be excited at the rough sidewalls and propagate along the *x*-direction for $\alpha=0$ and $\pi/2$. $\alpha$ is the angle between the nanowalls and the polarization direction of light.

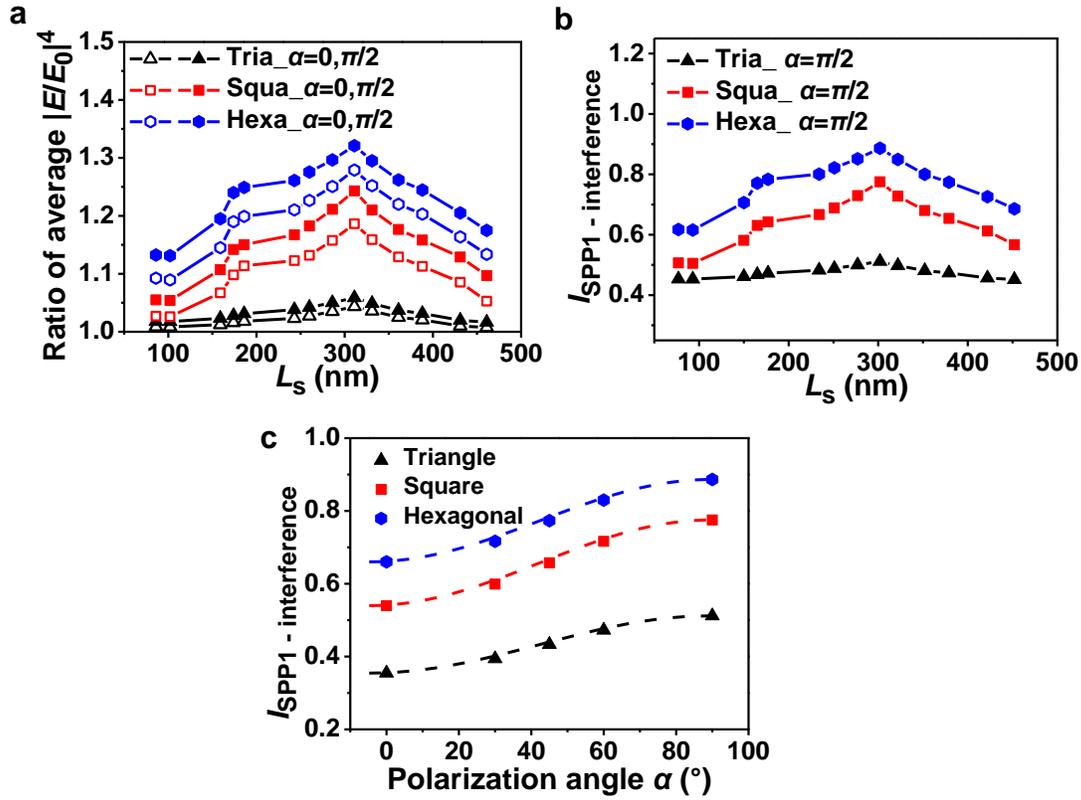

**Figure S9. The relationships between SPP1 wave interference effects with sidewall length and polarization angle $α$ calculated by FDTD calculations. a**, The ratio of the average of the fourth power of electric field intensities on sidewall surfaces of the rough triangular, square and hexagonal 36 nm Au/198 nm SiO$_2$ nanogrids to those of nanowalls for $α$ of 0 and $π/2$ with the increased sidewall length $L_s$. **b**, Calculated interference intensities of SPP1 wave excited at the rough sidewalls of triangular, square and hexagonal 36 nm Au/198 nm SiO$_2$ nanogrids with the polarization angles of $α = π/2$ against the increased sidewall length $L_s$. **c**, Calculated interference intensities of SPP1 wave taking place on a nanowall of triangular, square and hexagonal 36 nm Au/198 nm SiO$_2$ nanogrids with 302 nm sidewall length against polarization angle $α$, which can be well fitted with $(I_{1,π/2} - I_{1,0})\sin^2 α + I_{1,0}$ ($I_{1,π/2}$ and $I_{1,0}$ is interference intensity of SPP1 wave for $α = π/2$ and 0, respectively).

## S3. Optical standing wave effect of the incident light

The incident light with *x*-polarization and the part reflected by the bottom gold surface of the hybrid nanogrids can form optical standing wave,[9] whose electric field intensity $E_z$ shows the spatial distributions in the direction of height $z$ as follows

$$\begin{aligned} E_z(z) &= E_0(z) + E_{0r}(z) \\ &= E_0 e^{i(-kz-\omega t)} + \sqrt{R_z} E_0 e^{i(kz+\pi-\omega t)} \\ &= e^{i(-\omega t)}\left[(1-\sqrt{R_z})E_0\cos(kz) - i(1+\sqrt{R_z})E_0\sin(kz)\right] \end{aligned} \quad (7)$$

The light intensity in height is

$$I_z(z) = |E_z(z)|^2 = (1+R_z)E_0^2 - 2\sqrt{R_z}E_0^2\cos(2kz) \quad (8)$$

Thus, the normalized $I_z$ is

$$\overline{I_z} = \frac{1}{z-z_0}\int_{z_0}^{z} I_z(z) = (1+\sqrt{R_z})^2 E_0^2 - \frac{\sqrt{R_z}E_0^2}{kz}\sin(2kz) \quad (9)$$

where $R_z$, the reflectance of the incident light with *x*-polarization for square 36 nm Au/SiO$_2$ nanogrids with 200 nm grid length and a smooth 7 nm thick Au film on both sides of a sidewall, can be derived to be 0.47 from FDTD calculations. $z_0$, the coordinate position of the reflection plane, equals the thickness of Au film by defining the upper surface of silicon as the plane with $z=0$. Thus, we got the change of the normalized intensity of the optical standing wave with the ratio of height to the incident light wavelength, $z$-$z_0/\lambda$, which is shown in Figure S10.

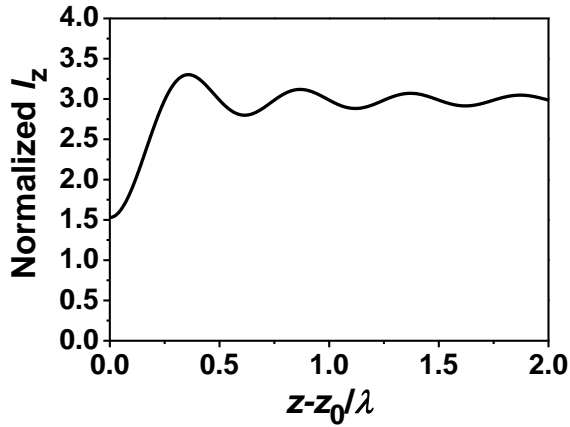

**Figure S10. Change of theoretical normalized intensity of the optical standing wave formed by the interference between the incident light with *x*-polarization and the reflected part against the ratio of height to the incident light wavelength, $z$-$z_0/\lambda$.** The oscillation behavior can be seen and the maximum interference intensity is observed to be around 0.36 (i.e. $z$-$z_0$ = 228 nm). With the $z$-$z_0/\lambda$ larger than 0.36, the oscillation tends to be weakened evidently.

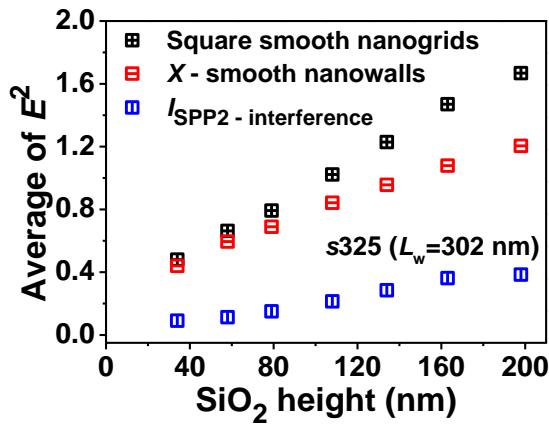

**Figure S11.** Calculated SiO$_2$ height dependences of the average squared electric field intensities on the surfaces (parallel to the *xz* plane) of smooth square 36 nm Au/SiO$_2$ nanogrids_*s*325 and *x*-nanowalls_*x*325 with 325 nm wall center distance, i.e. $L_w$ = 302 nm, and the corresponding SiO$_2$ height dependences of the average interference intensities of SPP2 wave for *s*325.

**Table S3.** The number *n* and center spacing *d* of, and the gap *g* between neighboring hemispheres / semiellipsoids Au nanoparticles for different $SiO_2$ heights and Au thicknesses derived from SEM observations.

| Au thickness (nm) | Parameters | SiO$_2$ height (nm) | | | | | | |
|---|---|---|---|---|---|---|---|---|
| | | 34 | 58 | 79 | 108 | 134 | 163 | 198 |
| 18 | *n* | 5 | 8 | 10 | 12 | 14 | 17 | 19 |
| | *d* (nm) | 6.3 | 7.1 | 7.9 | 9.0 | 9.5 | 9.9 | 10.4 |
| | *g* (nm) | -3.7 | -2.9 | -2.1 | -1.0 | -0.5 | 0.1 | 0.4 |
| 27 | *n* | 4 | 6 | 8 | 10 | 12 | 14 | 17 |
| | *d* (nm) | 7.7 | 8.7 | 9.7 | 10.8 | 11.2 | 11.6 | 11.8 |
| | *g* (nm) | -4.3 | -3.3 | -2.3 | -1.2 | -0.8 | -0.4 | -0.2 |
| 36 | *n* | 4 | 6 | 7 | 9 | 11 | 13 | 15 |
| | *d* (nm) | 8.6 | 9.4 | 10.3 | 11.4 | 11.9 | 12.6 | 12.8 |
| | *g* (nm) | -4.4 | -3.6 | -2.7 | -1.6 | -1.1 | -0.4 | -0.2 |

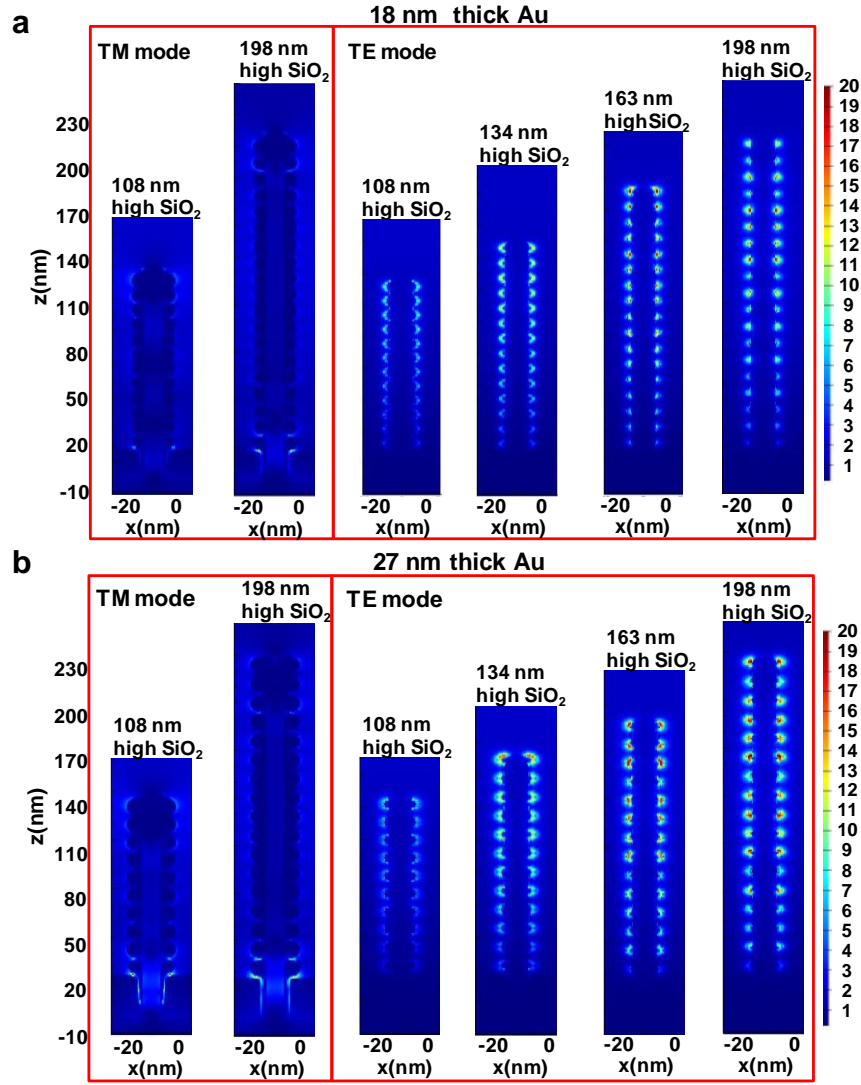

**Figure S12. Calculated spatial distributions of the electric field intensities on the cross sections parallel to the *xz* plane for 18 and 27 nm Au/SiO$_2$ *y*-nanowalls with 200 nm center distance *D* and different heights for TM and TE modes. a,** For 18 nm thick Au the maximum localized electric field intensity considering SPP1 coupling with LSPR2 is observed at the SiO$_2$ height of 163 nm for TE mode. Here, hemisphere-like Au nanoparticles with an average radius of 5 nm are taken for FDTD calculations. **b,** For 27 nm thick Au the change of the localized electric field with the increased height is similar to that for 36 nm thick Au, but with the lower values. Semiellipsoid-like Au nanoparticles with the average lengths of semi-principal axes, *a=b=* 6, and *c=* 6.5 nm, are employed for FDTD calculations.

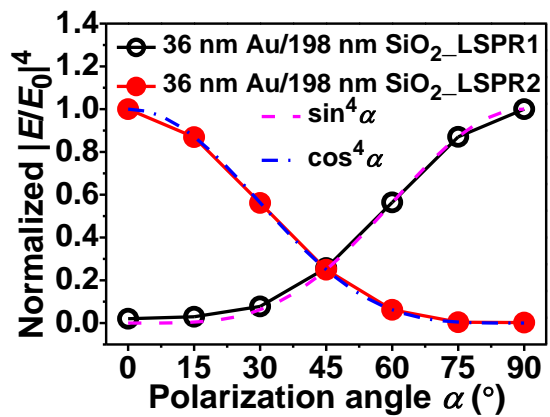

**Figure S13. The normalized $|E/E_0|^4$ for LSPR1 and LSPR2 of rough 36 nm Au/198 nm SiO$_2$ nanowalls with 151 nm sidewall spacing $L_w$ as a function of polarization angle $\alpha$ (solid line).** The relationships can be described as $\sin^4\alpha$ (dashed line) and $\cos^4\alpha$ (dashed dot line), respectively.

**S4. Coupling coefficients of different nanogrids**

To get the averaged $|E/E_0|^4$ for triangular, square and hexagonal nanogrids (Figure S14), the values of the average $|E/E_0|^4$ of LSPR1 and LSPR2 for TM and TE mode, respectively (Figure 3d) multiplied by their respective coupling coefficients are added together to be shown in Figure 4a-4c. These coupling coefficients were closely related to the symmetry and dimension of nanogrids and the polarization of incident light. The results by theoretical analyses are shown in Table S4 and Figure S15.

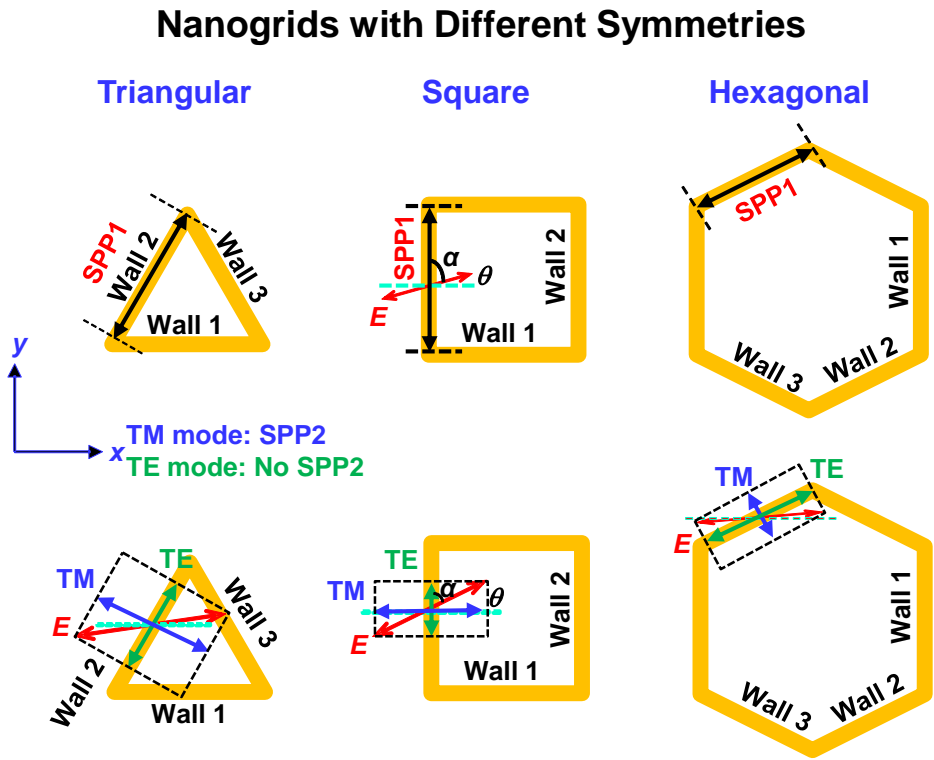

**Figure S14. Models of triangular, square and hexagonal nanogrids for calculations of the averaged $|E/E_0|^4$ with SPP wave coupling excitation effects considered.** Here, polarization angle $\theta$ and $\alpha$ is the angle between the polarization direction of light and *x*-direction, and between the nanowalls and the polarization direction of light, respectively.

**Table S4.** Polarization angle $\theta$ dependences of the average $|E/E_0|^4$ for triangular, square and hexagonal nanogrids.

| Triangular [a)] | | Wall 1 [a),b)] | Wall 2 [a),b)] | Wall 3 [a),b)] |
|---|---|---|---|---|
| Wall 1<br>$\alpha = -\theta$ | TM mode | $E_{LSPR1}^4\sin^4(-\theta)\,(I_{11}+1)$ | | |
| | TE mode | $E_{LSPR2}^4\cos^4(-\theta)\,(I_{11}+1)$ | | |
| Wall 2<br>$\alpha = \pi/3-\theta$ | TM mode | | $E_{LSPR1}^4\sin^4(\pi/3-\theta)\,(I_{12}+1)$ | |
| | TE mode | | $E_{LSPR2}^4\cos^4(\pi/3-\theta)\,(I_{12}+1)$ | |
| Wall 3<br>$\alpha = -\pi/3-\theta$ | TM mode | | | $E_{LSPR1}^4\sin^4(-\pi/3-\theta)(I_{13}+1)$ |
| | TE mode | | | $E_{LSPR2}^4\cos^4(-\pi/3-\theta)(I_{13}+1)$ |
| **Square** | | **Wall 1** | **Wall 2** | |
| Wall 1<br>$\alpha = -\theta$ | TM mode | $E_{LSPR1}^4\sin^4(-\theta)\,(I_{11}+1)$ | $E_{LSPR1}^4\sin^4(\pi/2-\theta)$<br>$\cdot(I_{12}+1)I_2\cos^2(\pi/2-\theta)$<br>$E_{LSPR2}^4\cos^4(\pi/2-\theta)$<br>$\cdot(I_{12}+1)I_2\cos^2(\pi/2-\theta)$ | |
| | TE mode | $E_{LSPR2}^4\cos^4(-\theta)\,(I_{11}+1)$ | | |
| Wall 2<br>$\alpha = \pi/2-\theta$ | TM mode | $E_{LSPR1}^4\sin^4(-\theta)$<br>$\cdot(I_{11}+1)I_2\cos^2(\theta)$<br>$E_{LSPR2}^4\cos^4(-\theta)$<br>$\cdot(I_{11}+1)I_2\cos^2(\theta)$ | $E_{LSPR1}^4\sin^4(\pi/2-\theta)\,(I_{12}+1)$ | |
| | TE mode | | $E_{LSPR2}^4\cos^4(\pi/2-\theta)\,(I_{12}+1)$ | |
| **Hexagonal** | | **Wall 1** | **Wall 2** | **Wall 3** |
| Wall 1<br>$\alpha = \pi/2-\theta$ | TM mode | $E_{LSPR1}^4\sin^4(\pi/2-\theta)\,(I_{11}+1)$ | $E_{LSPR1}^4\sin^4(\pi/6-\theta)$<br>$\cdot(I_{12}+1)I_2\cos^2(\theta)$<br>$E_{LSPR2}^4\cos^4(\pi/6-\theta)$<br>$\cdot(I_{12}+1)I_2\cos^2(\theta)$ | $E_{LSPR1}^4\sin^4(-\pi/6-\theta)$<br>$\cdot(I_{13}+1)I_2\cos^2(\theta)$<br>$E_{LSPR2}^4\cos^4(-\pi/6-\theta)$<br>$\cdot(I_{13}+1)I_2\cos^2(\theta)$ |
| | TE mode | $E_{LSPR2}^4\cos^4(\pi/2-\theta)\,(I_{11}+1)$ | | |
| Wall 2<br>$\alpha = \pi/6-\theta$ | TM mode | $E_{LSPR1}^4\sin^4(\pi/2-\theta)$<br>$\cdot(I_{11}+1)I_2\cos^2(-\pi/3-\theta)$<br>$E_{LSPR2}^4\cos^4(\pi/2-\theta)$<br>$\cdot(I_{11}+1)I_2\cos^2(-\pi/3-\theta)$ | $E_{LSPR1}^4\sin^4(\pi/6-\theta)\,(I_{12}+1)$ | $E_{LSPR1}^4\sin^4(-\pi/6-\theta)$<br>$\cdot(I_{13}+1)I_2\cos^2(-\pi/3-\theta)$<br>$E_{LSPR2}^4\cos^4(-\pi/6-\theta)$<br>$\cdot(I_{13}+1)I_2\cos^2(-\pi/3-\theta)$ |
| | TE mode | | $E_{LSPR2}^4\cos^4(\pi/6-\theta)\,(I_{12}+1)$ | |
| Wall 3<br>$\alpha = -\pi/6-\theta$ | TM mode | $E_{LSPR1}^4\sin^4(\pi/2-\theta)$<br>$\cdot(I_{11}+1)I_2\cos^2(\pi/3-\theta)$<br>$E_{LSPR2}^4\cos^4(\pi/2-\theta)$<br>$\cdot(I_{11}+1)I_2\cos^2(\pi/3-\theta)$ | $E_{LSPR1}^4\sin^4(\pi/6-\theta)$<br>$\cdot(I_{12}+1)I_2\cos^2(\pi/3-\theta)$<br>$E_{LSPR2}^4\cos^4(\pi/6-\theta)$<br>$\cdot(I_{12}+1)I_2\cos^2(\pi/3-\theta)$ | $E_{LSPR1}^4\sin^4(-\pi/6-\theta)(I_{13}+1)$ |
| | TE mode | | | $E_{LSPR2}^4\cos^4(-\pi/6-\theta)(I_{13}+1)$ |

[a)] Polarization angles $\theta$ and $\alpha$ is the angle between the polarization direction of light and $x$-direction, and between the nanowalls and the polarization direction of light, respectively (see Figure S14);

[b)] $I_{1i}$ (i=1, 2 and 3) are the interference intensities of SPP1 wave for walls1, 2 and 3, respectively (see Figure 2b, S9b, S9c and S14) and $I_2$ is the interference intensity of SPP2 wave for TM mode shown in Figure 2e.

Based on Table S4 and the light with the polarization angle of $\theta$, the coupling coefficients of LSPR1 and LSPR2 of triangular, square and hexagonal nanogrids are derived to be

$$C_{\text{LSPR1-tria}\theta} = [\sin^4\theta(\Delta I_1 \sin^2(0-\theta) + I_{1,0} + 1)$$
$$+ \sin^4(\frac{\pi}{3} - \theta)(\Delta I_1 \sin^2(\frac{\pi}{3} - \theta) + I_{1,0} + 1) + \sin^4(\frac{-\pi}{3} - \theta)(\Delta I_1 \sin^2(\frac{-\pi}{3} - \theta) + I_{1,0} + 1)]/3$$
$$C_{\text{LSPR1-squa}\theta} = 2[\sin^4\theta(1 + I_2\cos^2\theta)(\Delta I_1 \sin^2(0-\theta) + I_{1,0} + 1)$$
$$+ \sin^4(\frac{\pi}{2} - \theta)(1 + I_2\cos^2(\frac{\pi}{2} - \theta))(\Delta I_1 \sin^2(\frac{\pi}{2} - \theta) + I_{1,0} + 1)]/4 \quad (10)$$
$$C_{\text{LSPR1-hexa}\theta} = 2[\sin^4(\frac{\pi}{2} - \theta)[1 + I_2(\cos^2(\frac{-\pi}{3} - \theta) + \cos^2(\frac{\pi}{3} - \theta))](\Delta I_1 \sin^2(\frac{\pi}{2} - \theta) +$$
$$I_{1,0} + 1) + \sin^4(\frac{\pi}{6} - \theta)[1 + I_2(\cos^2\theta + \cos^2(\frac{\pi}{3} - \theta))](\Delta I_1 \sin^2(\frac{\pi}{6} - \theta) + I_{1,0} + 1)$$
$$+ \sin^4(\frac{-\pi}{6} - \theta)[1 + I_2(\cos^2\theta + \cos^2(\frac{-\pi}{3} - \theta))](\Delta I_1 \sin^2(\frac{-\pi}{6} - \theta) + I_{1,0} + 1)]/6$$

and

$$C_{\text{LSPR2-tria}\theta} = [\cos^4\theta(\Delta I_1 \sin^2(0-\theta) + I_{1,0} + 1)$$
$$+ \cos^4(\frac{\pi}{3} - \theta)(\Delta I_1 \sin^2(\frac{\pi}{3} - \theta) + I_{1,0} + 1) + \cos^4(\frac{-\pi}{3} - \theta)(\Delta I_1 \sin^2(\frac{-\pi}{3} - \theta) + I_{1,0} + 1)]/3$$
$$C_{\text{LSPR2-squa}\theta} = 2[\cos^4\theta(1 + I_2\cos^2\theta)(\Delta I_1 \sin^2(0-\theta) + I_{1,0} + 1)$$
$$+ \cos^4(\frac{\pi}{2} - \theta)(1 + I_2\cos^2(\frac{\pi}{2} - \theta))(\Delta I_1 \sin^2(\frac{\pi}{2} - \theta) + I_{1,0} + 1)]/4 \quad (11)$$
$$C_{\text{LSPR2-hexa}\theta} = 2[\cos^4(\frac{\pi}{2} - \theta)[1 + I_2(\cos^2(\frac{-\pi}{3} - \theta) + \cos^2(\frac{\pi}{3} - \theta))](\Delta I_1 \sin^2(\frac{\pi}{2} - \theta) +$$
$$I_{1,0} + 1) + \cos^4(\frac{\pi}{6} - \theta)[1 + I_2(\cos^2\theta + \cos^2(\frac{\pi}{3} - \theta))](\Delta I_1 \sin^2(\frac{\pi}{6} - \theta) + I_{1,0} + 1)$$
$$+ \cos^4(\frac{-\pi}{6} - \theta)[1 + I_2(\cos^2\theta + \cos^2(\frac{-\pi}{3} - \theta))](\Delta I_1 \sin^2(\frac{-\pi}{6} - \theta) + I_{1,0} + 1)]/6$$

Here, $\Delta I_1 = I_{1,\alpha=\pi/2} - I_{1,\alpha=0}$ and $I_{1,0} = I_{1,\alpha=0}$, in which $I_{1,\alpha=0}$ and $I_{1,\alpha=\pi/2}$ are the interference intensities of SPP1 wave for $\alpha=0$ and $\pi/2$, respectively (Figure 2b and Figure S9b), and $I_2$ is the interference intensity of SPP2 wave for TM mode (Figure 2e).

Then, we can get the averaged $|E/E_0|^4$ of these nanogrids using the following formula

$$\overline{|E_\theta/E_0|^4} = C_{\text{LSPR1-pattern}\theta}\overline{|E_{\text{LSPR1},\alpha=\pi/2}/E_0|^4} + C_{\text{LSPR2-pattern}\theta}\overline{|E_{\text{LSPR2},\alpha=0}/E_0|^4} \quad (12)$$

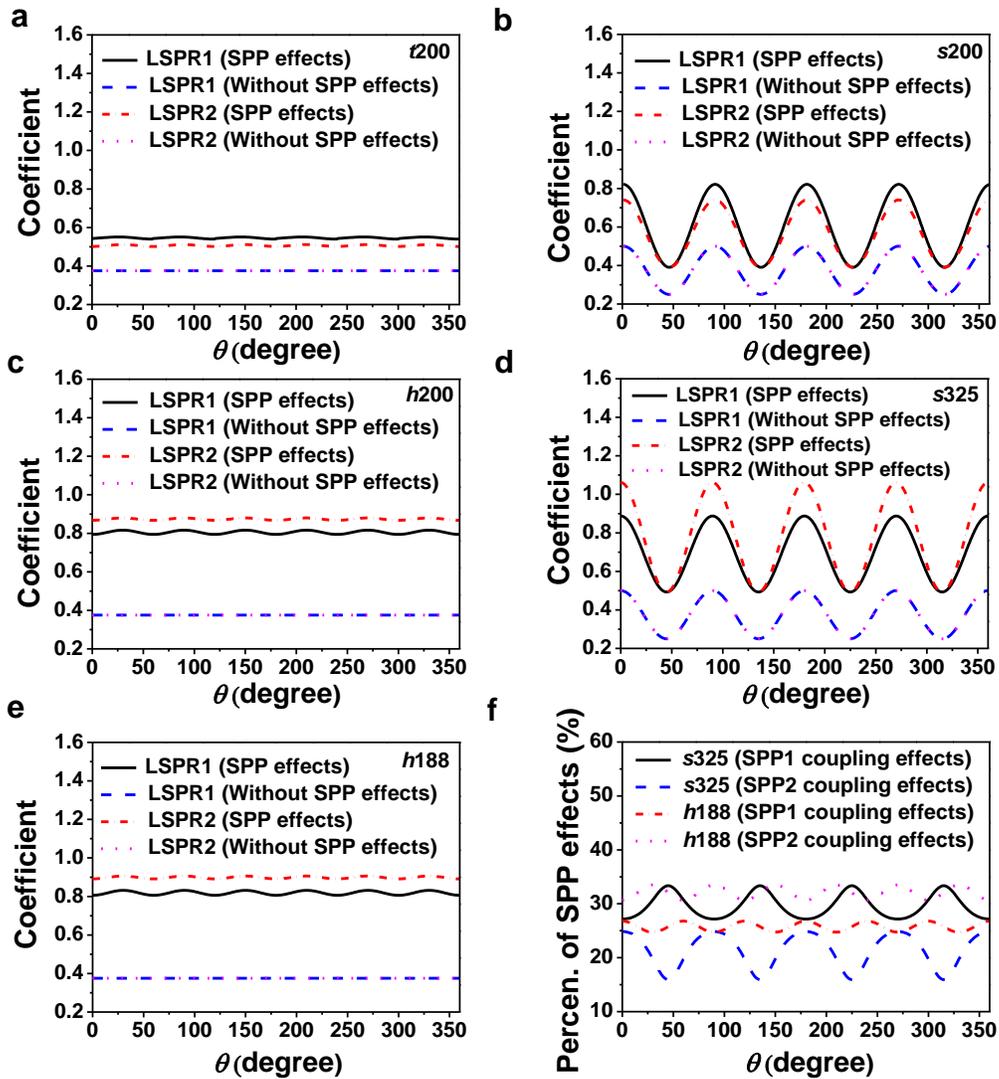

**Figure S15. Polarization angle $\theta$ dependences of the coupling coefficients of different nanogrids and the corresponding contribution percentages of SPP wave coupling excitation effects. a-e,** Polarization angle $\theta$ dependences of the coupling coefficients $C_{LSPR1}$ and $C_{LSPR2}$ of triangular, square and hexagonal nanogrids with and without the SPP wave coupling excitation effects considered. **f,** Percentages of intrinsic electric field enhancement derived from the SPP1 and SPP2 wave coupling excitation effects for square ($s$325) and hexagonal ($h$188) 36 nm Au/198 nm SiO$_2$ nanogrids versus polarization angle $\theta$.

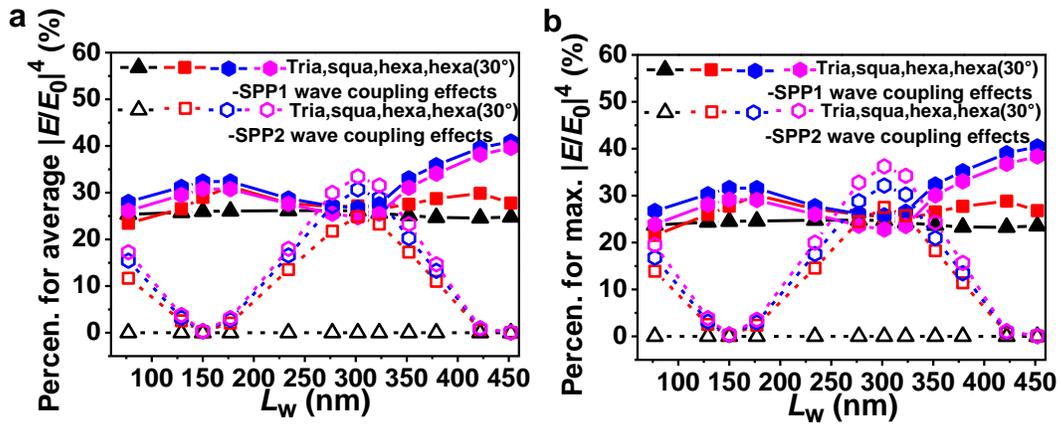

**Figure S16. Sidewall spacing $L_w$ dependences of the contributions of SPP1 and SPP2 wave coupling excitation effects to the average and maximum $|E/E_0|^4$ of different 36 nm Au/198 nm SiO$_2$ nanogrids, respectively. a,** The contributions of SPP1 and SPP2 wave coupling excitation effects to the average $|E/E_0|^4$. **b,** The contributions of SPP1 and SPP2 wave coupling excitation effects to the maximum $|E/E_0|^4$.

**Table S5.** Comparison of theoretical results of multiple effects of triangular, square and hexagonal 3D Au/SiO$_2$ periodic nanogrids with the heights of 34 and 198 nm and various dimensions.

| | | Theoretical results | | | | | |
|---|---|---|---|---|---|---|---|
| | | **34** | | **198** | | | |
| | | Height of SiO$_2$ (nm) | | | | | |
| **Multiple different effects (Influencing factors)** | | **s200** | **t200** | **s200** | **h200** | **s325*[a]** | **h188*[a]** |
| | | Symmetry & grid length $L_p$ (nm) | | | | | |
| | | 177 | 150 | 177 | 323 | 302 | 302 |
| | | Sidewall spacing $L_w$ (nm) | | | | | |
| **Excitation sources of nanogrids** ($\|E/E_0\|^2$) | Optical standing wave (SiO$_2$ height) | 0.53 | | 1.17 | | | |
| | SPP1 interference (Symmetry, sidewall length and polarization) | | 0.325 =[b] 0.477∥[b] | 0.446 = 0.647∥ | 0.557 = 0.785∥ | 0.529 = 0.768∥ | 0.556 = 0.784∥ |
| | SPP2 interference (TM mode) (Sidewall spacing and SiO$_2$ height) | 0.005 | 0 | 0.024 0.024 | 0.347 | 0.38 | 0.38 |
| **LSPR of nanowalls** ($\|E/E_0\|^4$) **(TM mode)** | Maximum LSPR ($\alpha = 0$) (×10$^6$) (Roughness and sidewall spacing) | 0.0055 | 3.40 | 3.96 | 5.06 | 5.02 | 5.02 |
| | Average LSPR ($\alpha = 0$) (×10$^5$) (Roughness and sidewall spacing) | 0.0012 | 0.676 | 0.80 | 1.19 | 1.18 | 1.18 |
| **Intrinsic electric field enhancement of nanogrids** | Average $\|E/E_0\|^4$ (×10$^5$) (Symmetry, sidewall spacing and polarization) | | 0.391 | 0.688 | 1.16 ($\theta=30°$) | 1.38 | 1.19 ($\theta=30°$) |
| | Average $\|E/E_0\|^4$ from SPP (%) (Symmetry, sidewall spacing and polarization) | | 26.0 | 33.4 | 57.0 ($\theta=30°$) | 52.0 | **58.2** ($\theta=30°$) |
| | Maximum $\|E/E_0\|^4$ (×10$^7$) (Symmetry, sidewall spacing and polarization) | | 0.451 | 0.586 | 1.20 ($\theta=30°$) | 1.07 | 1.22 ($\theta=30°$) |
| | Maximum $\|E/E_0\|^4$ from SPP (%) (Symmetry, sidewall spacing and polarization) | | 24.5 | 32.5 | 57.8 ($\theta=30°$) | 52.9 | **59.0** ($\theta=30°$) |

[a] * represents the optimized structures;
[b] = and ∥ represent $\alpha = 0$ and $\alpha = \pi/2$, respectively.

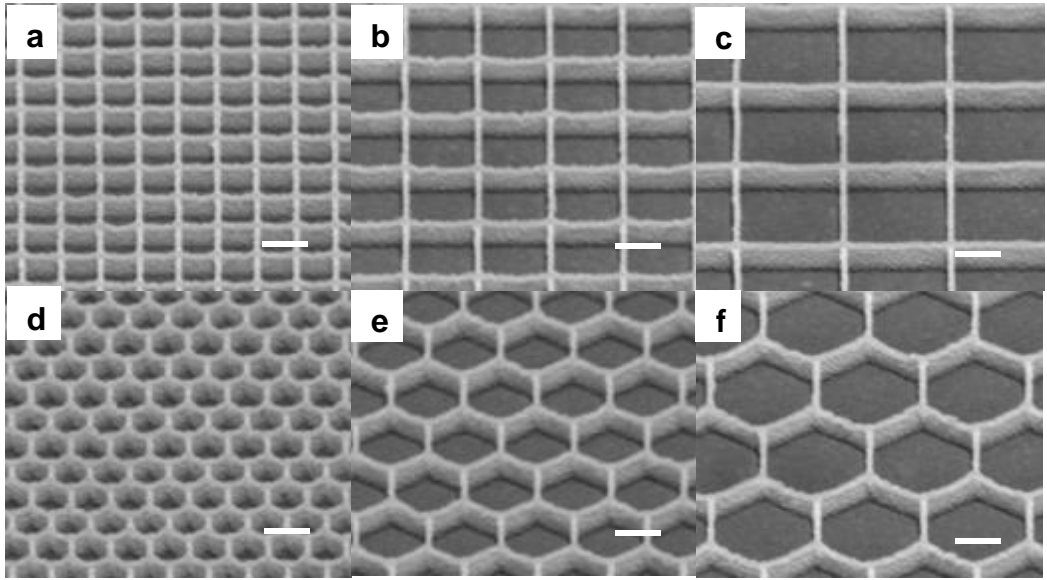

**Figure S17. Tilt SEM images of square and hexagonal 36 nm Au/198 nm SiO$_2$ nanogrids with different sidewall spacings. a-c,** Square nanogrids with 151, 302 and 452 nm sidewall spacing (i.e. *s*174, *s*325 and *s*475, respectively) with the sidewall width of 23 nm considered. **d-f,** Hexagonal nanogrids with the same sidewall spacings (i.e. *h*100, *h*188 and *h*274, respectively) with the sidewall width of 23 nm considered. Scale bars: 200 nm.

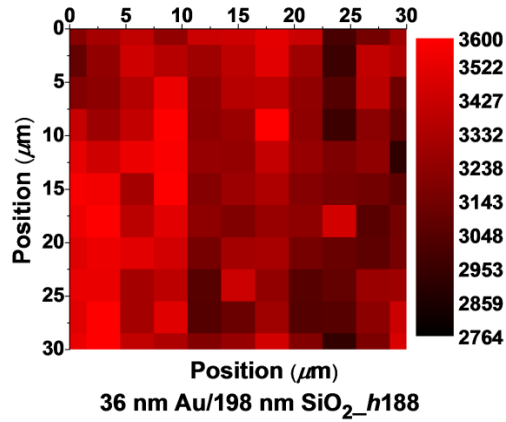

**Figure S18. Raman intensity mapping of R6G decorated hexagonal 36 nm Au/198 nm SiO$_2$ nanogrids_*h*188 as SERS probes.** 121 data across an area of 30μm × 30 μm were collected using a beam size of 1 μm at a step size of 3 μm upon SERS measurements. The nonuniformity is derived to be about 5.56%, which is quite good.

### S5. Calculation of SERS enhancement factor

The SERS enhancement factor (SERS EF) can be estimated by the following equation:[10]

$$\text{EF} = \frac{C_0 \times I_{\text{SERS}}}{I_0 \times C_{\text{SERS}}} \tag{13}$$

where $I_{\text{SERS}}$ and $I_0$ are the Raman scattering intensities of the peak at 1360 cm$^{-1}$ for R6G for hexagonal 36 nm Au/198 nm SiO$_2$ nanogrids_*h*188 with the polarization angle of 30° and 36 nm thick Au film, respectively, and $C_{\text{SERS}}$ and $C_0$ are the molar concentrations of R6G aqueous solution, 2.5×10$^{-11}$ and 10$^{-2}$ M used in this study, respectively, from which the SERS EF was derived to be 3.4×10$^8$. Therefore, the optimization of probes leads to a high SERS EF. The SERS EF calculated using the method in the manuscript better reflects the enhancement effect by the maximum electromagnetic field instead of the average enhancement effect due to the very low concentration molecules, $2.5\times10^{-11}$ M, which are probably decorated at those hot spots with the maximum or stronger electromagnetic field enhancement.